\documentclass[fleqn,useAMS,usenatbib]{mn2e}
\usepackage{graphicx}
\usepackage{amssymb}
\usepackage{amsmath}
\title[The REFLEX II power spectrum]{\textbf{The REFLEX II galaxy cluster survey:\\power spectrum analysis}}
\author[Balaguera-Antol\'{\i}nez et al.]{\parbox{\textwidth}{
A. Balaguera-Antol\'{\i}nez$^{1}$\thanks{E-mail: abalan@mpe.mpg.de}, 
Ariel G. S\'anchez$^{1}$, 
H. B\"ohringer$^{1}$,
C. Collins$^{2}$,
L. Guzzo$^{3}$, 
S. Phleps$^{1}$
}\\\\
$^{1}$Max Planck Institute f\"ur Extraterrestrische Physik, D-85748, Garching, Germany\\
$^{2}$Liverpool John Moores University, Liverpool, U.K.\\
$^{3}$INAF, Osservatorio Astronomico di Brera, Milano, Italy \\
}

\begin{document}
\pagerange{\pageref{firstpage}--\pageref{lastpage}} \pubyear{2010}
\maketitle

\label{firstpage}

\begin{abstract}
We present the power spectrum of galaxy clusters measured from the new ROSAT-ESO Flux-Limited X-Ray (REFLEX II) galaxy cluster
catalogue. This new sample extends the flux limit of the original REFLEX to $1.8 \times 10^{-12}\,{\rm erg}\,{\rm s}^{-1}\,{\rm cm}^{-2}$, yielding a total
of 911 clusters with  $\geq 94$ per cent completeness in redshift follow-up. The analysis of the data is improved by creating a set of $100$
REFLEX II-like mock galaxy cluster catalogues built from a suite of large volume $\Lambda$CDM $N$-body simulations (L-BASICC II). The measured power spectrum is in agreement with the predictions from a $\Lambda$CDM cosmological model. The measurements show the expected increase in the amplitude of the power spectrum with increasing X-ray luminosity. On large scales, we show that the shape of the measured power spectrum is compatible with a scale independent bias and provide a model for the amplitude that allows us to connect our measurements with a cosmological model. By implementing a luminosity-dependent power spectrum estimator, we observe that the power spectrum measured from the REFLEX II sample is weakly affected by flux-selection effects.  The shape of the measured power spectrum is compatible with a featureless power spectrum on scales $k>0.01\,h$ Mpc$^{-1}$ and hence no
statistically significant signal of baryonic acoustic oscillations can be detected. We show that the measured REFLEX II power spectrum
displays signatures of non-linear evolution. 
\end{abstract}


\section{Introduction}

Within the last decade, three foundational observational probes have been well
recognized as opening up observational windows to reveal some of the most valuable secrets of the Universe. These are: the temperature fluctuations of the cosmic microwave background radiation (CMB), recently measured with high
precision by the Wilkinson Microwave Anisotropy Probe (\emph{WMAP}) satellite \citep{spergel07, komatsu10}; the Hubble diagram inferred from
the  Type Ia supernovae (SNe) observations \citep[e.g.][]{perlmutter, riess}
and the measurement of the large scale structure (LSS) of the Universe as traced by the spatial distribution of galaxies
\citep{percival02,teg_sdss,sanchez06,percival_sdss,percival_last,reid}.

With the completion of large redshift surveys, such as the Two-degree Field Galaxy Redshift Survey
($2$dFGRS)\footnote{http://msowww.anu.edu.au/2dFGRS/} and  the Sloan Digital Sky Survey 
(SDSS)\footnote{http://www.sdss.org/} it has been possible to push the level of accuracy of LSS studies. 
The recent detection of the baryon acoustic oscillations (BAO, \citealp[e.g.][]{ei, cole2df,gaz,
ariel_last,percival_last}) in redshift surveys have been key to this progress.
Over the last few years these observations have established the concordance cosmological model, based on a flat space-time
in a current phase of accelerated expansion due to the presence of a dominating dark energy component, whose equation of
state is compatible with Einstein's cosmological constant $\Lambda$. This is the so-called $\Lambda$CDM cosmological model.

Galaxy cluster samples also provide valuable information to obtain constraints on cosmological parameters.
Galaxy clusters are the largest bounded structures in the Universe. They are associated with the highest peaks in the
matter density field and are recognized as biased tracers of the underlying matter distribution \citep[e.g.][]{bardeen}.
Their deep potential wells make them the largest astrophysical laboratories in the Universe, where the
combination of gravitation and baryonic physics has been intensively studied through the analysis of cluster properties
such as scaling relations \citep[e.g.][]{lm0,pratt,mantzI}, density profiles \citep[e.g.][]{makino}, pressure profiles
\citep[e.g.][]{pressure}, baryon fractions \citep[e.g.][]{baryon_frac}, etc. 
The abundances of galaxy clusters determined by their luminosity function \citep[e.g.][]{reflex_lf}
can also be used to constrain parameters like the matter content in the universe $\Omega_{\rm m}$, and the amplitude of
the density fluctuations characterized by $\sigma_{8}$, the rms linear perturbation
theory variance in spheres of radius $8\,$ Mpc $h^{-1}$ \citep[e.g.][]{reflex1_s8}. 
The spatial distribution of galaxy clusters, characterized by its power spectrum (or correlation function),
provides useful information about the cosmological model of the Universe.
The shape of this measurement is particularly sensitive to the parameter combination $\Omega_{\rm m}h$, which complements the
constraints on these parameters obtained from the analysis of fluctuations in the CMB. 
Furthermore, the amplitude of the galaxy cluster power spectrum also contains important information that can be 
related to theoretical models in a more direct way than for measurements based on galaxy samples \citep[e.g.][]{moscardini}.

\begin{figure*}
  \includegraphics[width=14cm]{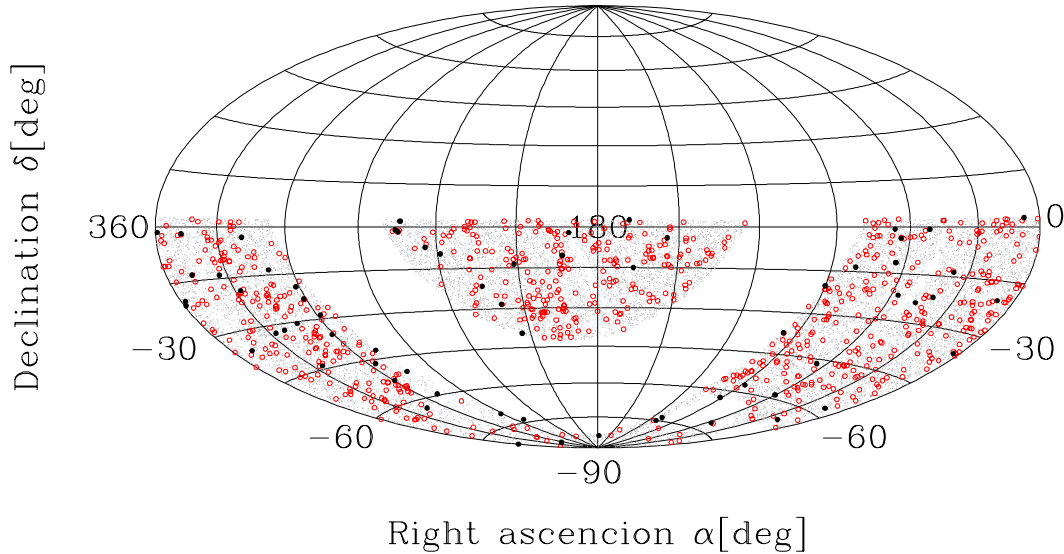}
  \caption[]{Distribution of REFLEX II clusters in equatorial coordinates. Open circles represent the position of the REFLEX II clusters. Filled circles represent the REFLEX II clusters without redshift. The points represent $5$ per cent of the random catalogue constructed with the REFLEX II  selection function. The empty regions in the Southern hemisphere corresponds to the cut in galactic coordinates $|b_{II}|<20^{o}$ of the Milky Way (band) and the Magellanic clouds. } \label{dsky}
\end{figure*}

In the past few years the ROSAT-ESO Flux Limited X-ray (REFLEX) catalogue \citep{hb_catalogo} has been used to measure fundamental
cosmological quantities. The REFLEX catalogue is based on \emph{ROSAT} All Sky Survey (RASS) observations \citep{truemper}, 
complemented with follow-up observations as described by \citet{guzzo_optical}, yielding spectroscopic redshifts for 447 clusters 
with flux limit of $3 \times 10^{-12}\, {\rm erg}\,{\rm s}^{-1}{\rm cm}^{-2}$ (in the \emph{ROSAT} energy band $0.1-2.4$\,keV).
The REFLEX catalogue was, to date, the largest statistically complete X-ray detected cluster sample.
The clustering properties of this survey were analyzed by means of the power spectrum \citep{reflex1}, the cluster correlation function
\citep{collins_cf}, cluster-galaxy cross-correlation functions \citep{ariel_reflexI} and Minkowski functionals \citep{kerscher}. 
Sub-samples of the REFLEX  catalogue complemented by detailed follow-up observations have been used to
constrain cluster scaling relations \citep[e.g.][]{lm0,stanek,pratt,mantzI}. 

In this paper we present the analysis of the power spectrum of the new REFLEX II catalogue.
The REFLEX II is an extension of the REFLEX catalogue to a lower limiting flux 
($1.8 \times 10^{-12}\, {\rm erg}\,{\rm s}^{-1}\,{\rm cm}^{-2}$) allowing the
inclusion of $464$ new clusters over the original sample and yielding a total of 911 clusters with spectroscopic redshifts for
$\sim 95$ per cent of the sample.  

In addition to the enlarged sample size of REFLEX II, several
improvements were made to the data reduction: (i) we use the
RASS survey product RASS III which gives a few percent more sky
exposure in formerly underexposed areas due to improved
attitude solutions, consequently recovering a few more
clusters at higher flux, (ii) for the
count rate to flux conversion an estimated temperature has
to be applied, which is now obtained with up-to-date scaling
relations based on the REXCESS Survey \citep{bohringer_07}
with the L-T relation described in \citet{pratt}; (iii) the
total flux and X-ray luminosity is now estimated inside the
radii of $r_{500}$ and $r_{200}$ (based on relations described in \citealt{pratt} and \citealt{arnaud_05}). These calculations now involve less
extrapolation than the  estimates for the previously used fiducial
radius.

Besides the advantages provided by a larger cluster sample, the power spectrum analysis presented here represents
an improvement over that of \citet{reflex1} in a number of ways. In particular, our analysis is complemented with a
set of $N$-body simulations, the L-BASICC II \citep{angulo,ariel1}, from which we constructed a suit of $100$ REFLEX II
mock catalogues. These catalogues were calibrated to reproduce the measured REFLEX II X-ray luminosity function.
Selection criteria of the REFLEX II sample were applied in
their construction, yielding a large suit of mocks that can be used to analyze the statistical methods applied
to the data. The details of the construction of these mock catalogues will be described in a
forthcoming paper (S\'anchez et. al. in preparation).

   Our ensemble of mock catalogues allowed us to show that it is possible to construct an accurate model 
of the shape and amplitude of the REFLEX II power spectrum.
This model includes the effects of the non-linear evolution of density fluctuations, redshift-space distortions
and halo bias, which introduce deviations in the clustering
signal with respect to the simple predictions of linear perturbation theory.
This will allow us to use the full information contained in the REFLEX II power spectrum to 
obtain constraints on cosmological parameters.

This paper is organized as follows. In Section \ref{sec_sel} we describe the REFLEX II sample and the survey selection function, followed by a
brief description of the construction of the mock catalogues in Section \ref{sec_mocks}. In Section \ref{sec_power} we describe the power spectrum estimator and show the measurements of the REFLEX II window function and the covariance matrix. In Section~\ref{lumi_bias} we model the amplitude of the power spectrum measured from the mock catalogues. In Section ~\ref{sec_malm} we explore the sensitivity of the REFLEX II sample to distortions induced by flux-selection effects. In Section \ref{sec_shape} we model the shape of the power spectrum. The model of the shape and the amplitude is applied to the REFLEX II sample in Section
~\ref{sec_q}. We end with our conclusions in Section \ref{sec_conc}.

Our fiducial cosmological model consist of a flat $\Lambda$CDM Universe with a matter energy density parameter of $\Omega_{\rm m}=0.25$, a dark
energy equation of state $w=-1$, a dimensionless Hubble parameter $h=0.7$ \footnote{The Hubble constant $H_{0}$ in units of
$100 \,{\rm km}\,{\rm s}^{-1}{\rm Mpc}^{-1}$.} and a spectral index of primordial scalar fluctuations $n_{s}=1$. Throughout this paper we always refer to the X-ray luminosity in the \emph{ROSAT} hard energy band
$0.1-2.4\, {\rm keV}$ and, whenever it is not explicitly written, its units are given in $10^{44}{\rm erg}\,{\rm s}^{-1}h^{-2}$.

\section{The REFLEX II Galaxy Cluster catalogue}\label{sec_sel}

\subsection{Catalogue construction}

Fig.~\ref{dsky} shows the angular positions of the 911 galaxy clusters of the REFLEX II sample. 
The catalogue covers $13924$\,deg$^{2}$ ($4.24$ sr) in the southern hemisphere ($\delta<2.5^{o}$) where
the Milky Way and the Magellanic Clouds have been excised to avoid
contamination due to stars and regions with high X-ray absorbing neutral hydrogen column density and high extinction
in the optical band. The limiting flux of the REFLEX II sample is $1.8\times 10^{-12}\,{\rm erg}\,\,{\rm s}^{-1}\,{\rm cm}^{-2}$ and includes 
$464$ new clusters in addition to the original REFLEX sample, which had $447$ clusters to a limiting flux of
$3\times 10^{-12}\,{\rm erg}\,\,{\rm s}^{-1}\,{\rm cm}^{-2}$. The detection technique is the same as that developed for the REFLEX
sample \citep{hb_catalogo, guzzo_optical}, with spectroscopic redshifts for $860$ clusters and spanning luminosities in
the range  $4.9\times 10^{40}\leq L_{X}/({\rm erg}\,\,{\rm s}^{-1}h^{-2})\leq 1.96\times 10^{45}$. The missing $\approx 6$ per cent of the redshifts will be obtained in observations made at La Silla in fall of 2010 and spring 2011. The angular distribution of X-ray clusters without redshift in the REFLEX II sample is also shown in Fig.\ref{dsky} (filled circles), displaying no particular pattern on the sky. We estimate that our
results will not change substantially due to this incompleteness. 

While a more detailed description of the sample construction and the derivation of the cluster parameters will be given
in a forthcoming paper (B\"ohringer et al. in preparation), we provide here a brief description of these measurements
and calculations. Source counts for the galaxy clusters in the RASS have been determined in the $0.5$ to $2$ keV energy
band by means of the growth curve analysis method described in \citet{hb_cga}. The growth curve method is tailored to
maximize the aperture in which the source counts are determined. The count rate (obtained by reference to the exposure
maps of the RASS) is then converted to a nominal flux, $F_n$, by means of XSPEC \citep{xspec} assuming a MEKAL plasma
model for a cluster temperature of 5 keV, a redshift of $z = 0$, a metallicity of $0.3$ solar and an value for the
hydrogen column density taken from the measurements of \citet{dickey}. The flux limit is imposed on the cluster
sample using this value of $F_n$. For clusters whose redshifts are known, an improved flux value, $F_X$, is determined
by recalculating the count rate to flux conversion for a temperature estimated via the X-ray
luminosity temperature relation as given by \citet{pratt} and by including the
proper band corrections (analogous to the optical K-correction) for the actual cluster redshift. The measured
flux is converted to an estimated flux within an aperture radius of $r_{500}$ (the radius in which the mean matter
density of the cluster is 500 times the critical density of the Universe) by means of relations given in \citet{pratt}.
The flux extrapolation (and in some cases interpolation) is achieved by assuming a cluster surface
brightness following a $\beta$-model \citep{beta_model} with $\beta = 2/3$ and core radius $r_c = (1/7)r_{500}$.
The cluster rest frame X-ray luminosity is calculated by means of the luminosity distance from $F_X$ by taking the
proper band corrections for the redshift into account.

For the determination of the sky-position dependent selection function, the minimum luminosity as a function of redshift and the position
on the sky $L_X^{\rm lim} (\alpha , \delta, z)$ is calculated assuming the nominal flux limit and taking into account
that for $5.4$ per cent of the sky the exposure is too short ($<100$ sec. mostly due to instrument shut-down during passages
of the radiation belts in South Atlantic Anomaly) to reach the nominal flux limit.
The values of $L_X^{\rm lim} (\alpha,\delta,z)$ are then derived by accounting for the proper
$F_X$ for given redshift and by performing an iterative backward engineering
of the above described process. The surveyed area defining the REFLEX II sample has been divided into 
$N_{\rm pix}=13902$ pixels with area $\approx 1\,{\rm deg}^{2}$. For each pixel centered on
equatorial coordinates $(\alpha_{i},\delta_{i})$ the limiting luminosity 
$L_X^{\rm lim} (\alpha_{i},\delta_{i},z)$ was tabulated in the range $0\leq z \leq 0.8$.
Given the minimum count rate of $20$ cts and the geometrical boundaries of the survey (see table $1$ of \citet{hb_catalogo}) we end up with a total of $787$ galaxy clusters.

 \begin{table}
\center
\begin{tabular}{c|c|c|c|c|c} \hline\hline
Sample $i$ &$L^{\rm min}_i$  &  $N_i$ & $\bar{z}_i$  & $N_{VLS}$ & $\bar{n}_{\rm VLS}$\\
&&\\
1&$0.015$ &   $661$ & $0.070$& $32 $ &$4.318$\\
2&$0.049$   & $625$ & $0.080$& $90$ &$1.856$ \\
3&$0.154$  & $ 512$& $0.099$& $150$  &$0.6562$ \\
4&$0.245$   & $441$& $0.112$ & $159$ &$0.3905$\\
5&$0.490 $  & $306$ & $0.136$ & $154$&$0.1535$\\
6&$0.588$   & $260$ & $0.143$& $157$ & $0.1155$ \\\hline\hline
\end{tabular}\caption{Minimum X-ray luminosities $L^{\rm min}_{i}$ ($i=1,\cdots,6$) used to define sub-samples of the REFLEX II catalogue, with a maximum redshift $z_{\rm max}=0.22$.
$N_i$ denotes the number of clusters in each sub-sample and $\bar{z}_{i}$ is the corresponding mean redshift. $N_{VLS}$ is the number of REFLEX II clusters and $\bar{n}_{VLS}$ the mean density in each sample volume-limited sample, in units of $10^{-6}({\rm Mpc}\,h^{-1})^{-3}$. X-ray luminosity in units $10^{44}{\rm erg}\,\,{\rm s}^{-1}\,h^{-2}$ in the energy band $0.1-2.4$ keV.}.
\label{tableVLS}
\end{table}

\subsection{The REFLEX II Selection Function}\label{lfun}
\begin{figure}
  \includegraphics[width=8.2cm, angle=0]{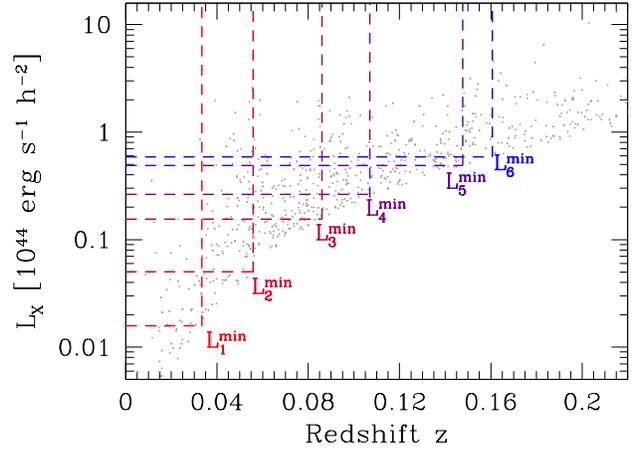}
  \caption[]{Luminosity-redshift diagram for the REFLEX II sample (points). Dashed lines schematically represents the volume-limited samples defined by the limiting luminosities of Table~\ref{tableVLS}. }\label{lz_vls}
\end{figure}

We measured the  REFLEX II X-ray luminosity function using the $V_{\rm max}$ estimator. We fitted the measurements by means of what we call an
\emph{extended Schechter function}, 
\begin{equation}\label{ex-sh}
\Phi(L_{X}){\rm d} L_{X}=n_0\left(\frac{L_{X}}{L_{\star}}\right)^{\alpha}{\rm e}_{q}(x) \frac{{\rm d} L_{X}}{L_{\star}},
\end{equation}
where ${\rm e}_{q}(x)$ denotes the so-called $q-$exponential function \citep[e.g.][]{tsallis} defined as 
\begin{equation}
{\rm e}_{q}(x)\equiv 
\begin{cases}
e^{x}& q=1, \\
(1+x(1-q))^{1/(1-q)}&  q\neq 1.\\
\end{cases}
\end{equation}
This parameterization provides a better description of the high luminosity tail of the X-ray luminosity function (S\'anchez et al., in prep).
We implemented a Markov Chain Monte Carlo (MCMC) technique to determine the best fit parameters ($n_0,\alpha,L_{\star},q$)
given by 
\begin{eqnarray}
n_0&=&(4.081^{+0.012}_{-0.010})\times 10^{-6}({\rm Mpc}\, h^{-1})^{-3}, \nonumber\\
\alpha&=&-1.536^{+0.006}_{-0.013},\nonumber\\ 
 L_{\star}&=&(0.6283^{+0.031}_{-0.028})\times 10 ^{44}{\rm erg}\,\,{\rm s}^{-1}\,h^{-2},\hspace{0.1cm}{\rm and} \nonumber\\
 q&=&1.313^{+0.010}_{-0.013}.
\end{eqnarray}
In order to explore the behavior of the clustering strength as a function of X-ray luminosity, we split the REFLEX II sample in six sub-samples characterized by a minimum luminosity $L^{\rm min}_{i}$. These are listed in Table \ref{tableVLS}, together their corresponding number of clusters $N_{i}$ and mean redshift $\bar{z}_{i}$ . For a given luminosity cut $L^{\rm min}_{i}$, the expected number density at
a position $\textbf{r}$, characterized by angular coordinates $(\alpha,\delta)$ and redshift $z$,
is obtained via integration of the X-ray luminosity function as
\begin{equation} \label{nbar}
  \bar{n}(\textbf{r};L^{\rm min}_{i})=\int_{\hat{L}(\textbf{r})}^{\infty} \, \Phi(L_{X})\,{\rm d} L_{X},
\end{equation}
where the lower integration limit $\hat{L}(\textbf{r})$ is given by the REFLEX II sensitivity map as
\begin{equation} \label{lhat}
  \hat{L}(\textbf{r})\equiv {\rm Max}(L^{\rm min}_{i}, L^{\rm lim}_{X}(\alpha,\delta,z)).
\end{equation}
We also constructed six volume-limited samples (hereafter VLS) using the same minimum luminosities. 
These VLS are schematically represented in Fig.~\ref{lz_vls}. 
Table \ref{tableVLS} also lists the number of clusters, the mean redshift and the cluster number density
for these sub-samples.

Finally, we created a random catalogue of $N_{\rm r}=2\times 10^{6}$ objects with luminosities greater
than $L_{X}=1.4\times 10^{43}{\rm erg}\,\,{\rm s}^{-1}\,h^{-2}$. In Fig.~\ref{dsky} we show the angular positions of a subset of
the random sample. Note the variations in the angular distribution of the random catalogue, which follow the
fluctuations in the sensitivity map of the REFLEX II survey.

\subsection{The mock catalogues}\label{sec_mocks}

In this section we give a short outline of the construction of a suit of REFLEX II mock catalogues. 
A more detailed description and analysis will be given in a forthcoming paper (S\'anchez et al. in prep).

We used the $z=0$ output of the 50 realizations from the L-BASICC II $N$-body simulations \citep{angulo,ariel1}
to generate a suit of REFLEX II mock catalogues. The cosmological model adopted in the simulations consists of a flat
$\Lambda$CDM universe with a matter energy-density parameter $\Omega_{\rm m}=0.237$, a baryon energy-density parameter of
$\Omega_{\rm b}=0.046$, a dimensionless Hubble parameter $h=0.73$, a dark energy equation of state $w=-1$, and an initial
matter power spectrum characterized by a scalar spectral index $n_{\rm s} = 0.954$ and normalized to
$\sigma_{8} = 0.77$,
which is in close agreement with the latest constraints on cosmological parameters from CMB and LSS measurements.
\citep[e.g.][]{sanchez06, spergel07, ariel_last, komatsu10}.
The low value of $\sigma_{8}$ is particularly important in order to obtain cluster number 
counts in accordance with observations.

Each of the L-BASICC II simulations follows the dark matter distribution using $448^3$ particles
over a comoving box of side $1.34\,{\rm Gpc}\,h^{-1}$. We used halo catalogues identified via a friend-of-friend 
(FoF) algorithm with a linking length of $b=0.2$. The resulting halo mass resolution is $M=1.75\times 10^{13}M_{\odot}h^{-1}$.

We assigned luminosities to the dark matter haloes in the simulations using a mass-luminosity $M-L_{X}$ 
relation with an intrinsic scatter $\sigma_{\ln L}=0.26$. This value was measured by \citet{stanek} under the assumption of
a flat cosmological model with $\Omega_{\rm m}=0.24$, close to our fiducial cosmology. Flux errors are included in
the luminosity assignment by assuming a fixed error in $L_{X}$ of $\delta L_{X}/L_{X}=20$ per cent, 
which characterizes the flux errors in the REFLEX II sample fairly well. For the mean $M-L_{X}$ relation we assumed a power law with a 
mass-dependent slope. Setting 
\begin{equation}
\ell =\log_{10}\left( \frac{\bar{L}_{X}}{10^{44}{\rm erg}\,\,{\rm s}^{-1}\,h^{-2}}\right),\,\,m=\log_{10}\left( \frac{M}{10^{14}M_{\odot}\,h^{-1}} \right),
\label{eq:teo_mass_lum}
\end{equation}
our mass-luminosity relation reads as
\begin{equation}\label{lum}
\ell=a+bm+cm^{2}.
\end{equation}
Using a MCMC technique, we calibrated the set of coefficients $(a,b,c)$ such that the resulting luminosity distribution of the illuminated 
dark matter haloes match the measured X-ray luminosity function from the REFLEX II sample. The best fit values are $a=-1.3164$, $b=1.8769$ and $c=-0.2955$. Changes below $1$ per cent in 
these parameters are observed when the flux error is reduced to $10$ per cent, although this value underestimates the observed flux errors of the 
REFLEX II sample. The goodness of this self-calibration is confirmed by comparing the REFLEX II luminosity function and the mean luminosity 
function determined from the mock catalogues, obtaining differences of $\leq 3$ per cent. Interestingly, this calibration of the $M-L_{X}$ 
relation shows a deviation from the power-law behavior at low luminosities. This is in agreement with recent observations and $N$-body
hydro-simulations \citep{puchwein,stanek_last}. Nevertheless we do not attempt to extract conclusions on the underlying mass-luminosity relation for a number of reasons. First, the FoF halo-finder algorithms overestimates the halo mass due to systematics introduced by the finite number of particles used to define a dark matter halo. \citet{warren} showed a way to correct for this systematics, which only depends on the FoF mass. For the lowest luminosity cut shown in Table~\ref{tableVLS}, the distribution of minimum halo masses selected by equation~(\ref{lum}) in the mocks peaks at $M\approx 5\times 10^{13}M_{\odot}\,h^{-1}$, which, following the correction of \citet{warren} corresponds to an offset of $\sim 40$ per cent with respect to unbiased mass estimations. Nevertheless, we do not attempt to correct the L-BASICC II FoF masses, for any correction would lead to a
new set of parameters $(a,b,c)$ which will still reproduce, by construction, the observed X-ray luminosity function. Secondly, there is not a clear one-to-one relation between FoF masses and the spherical overdensity masses \citep[e.g.][]{lukic10}, which are usually implemented in the calibration of cluster scaling relations \citep[e.g.][]{pratt}.

We next transformed the coordinates of the haloes to redshift-space using $r\to r+\textbf{v}\cdot\hat{\textbf{r}}/H_{0}$ where $\textbf{v}$ 
is the peculiar velocity of the center of mass of the haloes. We neglected spectroscopic redshift errors for being of the order of 
$\sigma\sim10^{-3}$ \citep{guzzo_optical}. Finally we observed the illuminated haloes through the REFLEX II mask and obtained a set of mock
 catalogues with the abundance and geometry in agreement with those of the REFLEX II sample.

We constructed two sets of mocks. One set consists of $50$ mocks covering the full volume of the REFLEX II sample (out to 
$z\approx 0.5$). For the second set, we noticed that the effective volume of the REFLEX II sample reaches a maximum
at redshift $z\approx 0.2$ (see Section  \ref{sec_cova}). Therefore, including clusters with higher redshift will not
help to improve the signal-to-noise ratio of the measurements. For this reason we set a maximum redshift of $z=0.22$ which allowed us to
construct $100$ independent mock catalogues out of the $50$ L-BASICC II realizations. Unless otherwise stated, we use the set of
$100$ mocks to carry out our statistical analysis. The set of $50$ mocks were used for consistency checks.

In the analysis of the REFLEX power spectrum, \citet{reflex1} constructed ten mock catalogues by 
assigning X-ray luminosities to the dark matter haloes in a small set of $N$-body simulations using  the $M-L_{X}$ relation of \citet{lm0}.
No intrinsic scatter around the mean $M-L_{X}$ relation, flux errors or missing flux corrections
were introduced in the mocks. Note that this difference with respect to our analysis will result in a different prediction
of the amplitude of the power spectrum for a given luminosity cut. In Section \ref{sec_amp} we address the modeling of the amplitude
of the power spectrum taking these details into account.

\section{Methods}\label{sec_power}

\subsection{The measurement of $P(k)$}\label{sec:method}

\begin{figure}
  \includegraphics[width=8cm, angle=0]{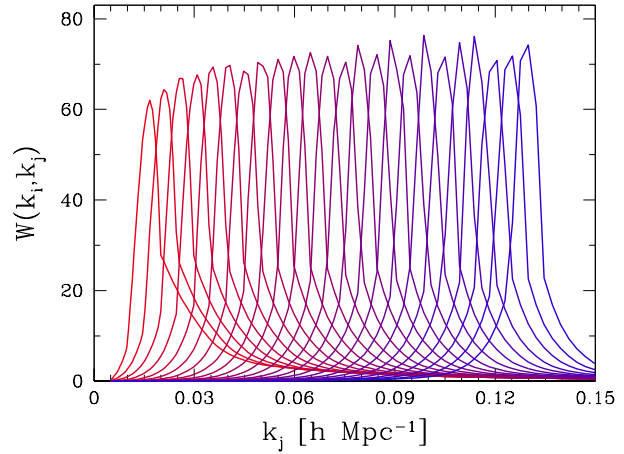}
\caption[]{Example of  the REFLEX II shell averaged window (matrix) function $W_{ij}$ for different modes $k_{i}$ determined 
for the sub-sample with limiting luminosity $L^{\rm min}_{2}$. Each curve is normalized to $\int {\rm d}k_{j} W(k_{i},k_{j})=1$. }\label{window_reflex}
\end{figure}

We have measured the power spectrum of the REFLEX II sample containing clusters with luminosities $L_{X}\geq L^{\rm min}_{1}$, which represents
$760$ objects in the redshift interval covered by the REFLEX II survey. We implemented the standard minimum variance weighting power spectrum
estimator  of \citealt{FKP} (hereafter FKP ), which defines a weighted density fluctuation 
\begin{equation}
F(\textbf{r})=w(\textbf{r})\left( n_{\rm c}(\textbf{r})-\alpha n_{\rm r}(\textbf{r})\right),
\end{equation}
where $n_{\rm c}(\textbf{r})$ ($n_{\rm r}(\textbf{r})$) is the number density of clusters in the real (random) catalogue for a given
luminosity cut (where we dropped the $L^{\rm min}$ dependence to avoid clutter). The parameter $\alpha$, given by 
\begin{equation}
\alpha=\frac{\int w(\textbf{r})n_{\rm c}(\textbf{r})\,{\rm d}^{3}r}{\int  w(\textbf{r})n_{\rm r}(\textbf{r})\,{\rm d}^{3} r },
\end{equation}
forces the fluctuation to have zero mean, $\int\,F(\textbf{r})\,{\rm d}^{3}r=0$. The optimal normalized weights $w(\textbf{r})$ are given by \citet{FKP} \begin{equation}\label{wop}
w(\textbf{r})=\left( \frac{1}{1+\bar{n}(\textbf{r})P_{\rm est}}\right) \left( \int \,\frac{\bar{n}(\textbf{r})}{1+\bar{n}(\textbf{r})P_{\rm est}}\,{\rm d}^{3}r \right)^{-1},
\end{equation}
 where $P_{\rm est}$ is an estimate of the amplitude of the power spectrum expected to be measured, for which we have chosen $P_{\rm est}=2\times 10^{4} ({\rm Mpc}\,h^{-1})^{3}$.
We used the FFTw algorithm \citep{fftw} embedding the REFLEX II volume in a cube divided in $N_{\rm grid}=512^{3}$ cells and implemented a
triangular shaped cloud mass assignment \citep{hockney} correcting afterwards for aliasing effects. The length of the sides of
the cube are determined by $L_{\rm box}=2r(z_{\rm max})$, with $z_{\rm max}=0.22$ which corresponds to a box size of $1.25$ Gpc $h^{-1}$ for our
fiducial cosmology. The fundamental mode is $\Delta k=2\pi/L_{\rm box}=0.0049\,h$ Mpc$^{-1}$. The Nyquist frequency for this box is
$k_{\rm Nyq}=1.27 h$ Mpc$^{-1}$, and we can ignore aliasing effects on wavenumbers smaller than $k\approx 0.7\,h$ Mpc$^{-1}$. We subtract the
shot noise and average in spherical shells to obtain the bin-averaged power spectrum $\hat{P}(k)$. This measurement is the
convolution of the underlying cluster power spectrum with $|W(\textbf{k})|^{2}$, the square of the Fourier transform of the REFLEX II
window function given by
\begin{equation}
W(\textbf{k})= \int  \bar{n}(\textbf{r})\,w(\textbf{r}){\rm e}^{-i \textbf{k}\cdot \textbf{r}}\,{\rm d}^{3}r\, . 
\end{equation}

\begin{figure}
  \includegraphics[width=8.2cm, angle=0]{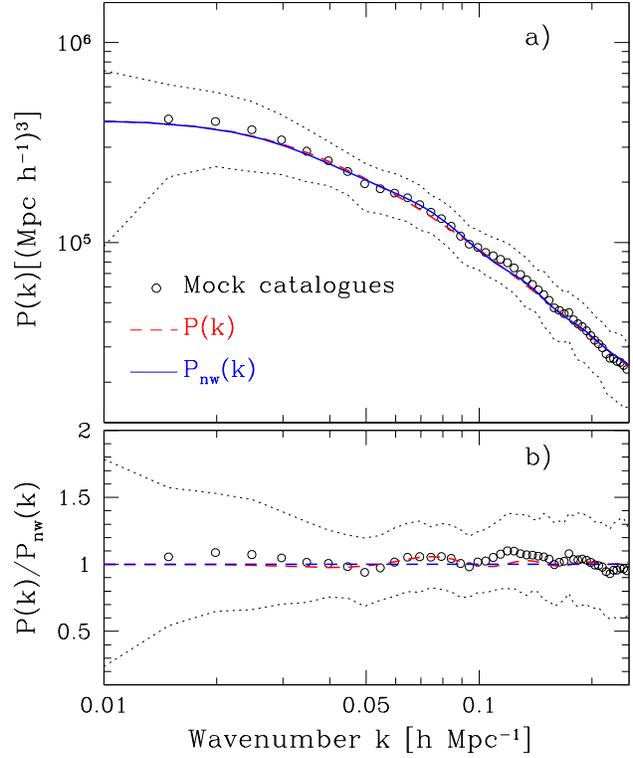}
  \caption[]{a) Mean power spectrum from the mock catalogues (open points) with its corresponding standard deviation (dotted lines), for the sub-sample characterized with the limiting luminosity $L^{\rm min}_{2}$. The solid line $P_{\rm nw}(k)$ represents a non-linear matter power spectrum $P(k)$ without BAO and convolved with the REFLEX II window function. The dashed line represents the same theoretical prediction with BAO, also convolved with the window function. b) Ratio of the mean mock power spectrum (open points), and the power spectrum $P(k)$ to the power spectrum $P_{\rm nw}(k)$. Dotted lines denotes the $1\sigma$ standard deviation.}\label{power_conv}
\end{figure}

We follow the procedure of \citet{cole2df} to construct the window 
function in matrix form $W_{ij}$, by using a Gauss-Legendre integration scheme \citep{numerical_rec}. 
The measured power spectrum can be written as a matrix multiplication 
\begin{equation}
\hat{P}(k_{i})=\sum_{j}W_{ij}P(k_{j})-W_{i 0},
\end{equation}
where $P(k_{j})$ is the underlying power spectrum and the term $W_{i 0}$ accounts for the integral constraint \citep{percival_sdss,reid}.
As an example, Fig.~\ref{window_reflex} shows some elements of the window matrix of the sample defined by the minimum luminosity
$L^{\rm min}_{2}$ (panel a) and the volume limited sample defined by the same luminosity cut (panel b). 
As expected, large scale modes receive contributions from intermediate and even small ($k\geq 0.3 h$ Mpc$^{-1}$) scales.

We used the window matrix to asses the possibility of detecting the signal of the BAO in the measured REFLEX II
power spectrum. The signature of BAO in the dark matter halo distribution of the L-BASICC II simulations has been analyzed both
in the spatial two-point correlation function \citep{ariel1} as well as in the power spectrum \citep{angulo}. We used the fitting formulae of \citet{hu} to compute the matter power spectrum for the cosmological model 
of the L-BASICC II simulations, including a non-linear correction computed with {\sc halofit} \citep{smith}.
We also computed a model with the same broad-band shape but without any baryonic oscillations, $P_{\rm nw}(k)$, provided as well by \citet{hu}.
Panel a) of Fig.~\ref{power_conv} shows the comparison of these theoretical models (solid and dashed lines), convolved with
the REFLEX II window function, with the mean power spectrum from the mock catalogues (open circles).
It can be seen that the difference between these two models is much smaller than the standard deviation in $P(k)$
that corresponds to the REFLEX II volume, which can be determined from the ensemble of mock catalogues
(dotted lines, see Section \ref{sec_cova}). This can be more clearly seen in panel b) of the same figure,
which shows the ratio of these power spectra to $P_{\rm nw}(k)$. The convolution with the window function damps
the acoustic oscillations in the power spectrum to a level where they cannot be distinguished from a model without
BAO. Furthermore, we computed the $\chi^{2}$ of these two models \citep[analytically marginalizing over the amplitude as 
described in][]{lewis} and found a difference of less than $\sim 3$ per cent between them.
We thus conclude that, due to the survey volume, no statistically significant signal of BAO can be detected in the
REFLEX II power spectrum.

A detection of BAO signatures in a galaxy cluster sample was claimed
by Miller et al. (2001) based on the measurements of the power spectrum of the Abell/ACO
cluster survey \citep{miller}.
This claim was based on different statistical methods from the ones
applied in this work and no mock catalogues were used to asses the statistical significance of
the detection. Our cluster sample contains approximately the same number of objects
as that used by Miller et al. (2001), while probing a larger volume.
Recently, \cite{hutsi_gc} reported a 2$\sigma$σ detection of BAO in the maxBCG
galaxy cluster survey, which probes a larger effective volume than the REFLEX II sample.

Finally, regarding the redshift incompleteness of the REFLEX II catalogue mentioned in Section 2.1, we verified that our
results are not substantially modified when the power spectrum is measured after randomly subtracting up to $20$ per cent
of the total number of clusters.

\begin{figure*}
\includegraphics[width=5.5cm, angle=0]{cova_contours_zcut_98mocks_lm26_color_e20.epsi}
\includegraphics[width=5.5cm, angle=0]{cova_contours_zcut_98mocks_lm26_color_e20_pvp.epsi}
\includegraphics[width=5.5cm, angle=0]{cova_contours_box_lmin26_z0_box_color.epsi}
\caption[]{Comparison of the correlation matrix $|r(k_{i},k_{j})|$ of the power spectra measured with a) the FKP estimator and b) the PVP 
estimator (see Section \ref{sec_malm}) for two luminosity-cuts. c) Correlation matrix determined 
from the illuminated dark matter halos of the L-BASICC II simulation with same luminosity-cuts.}\label{rc_2dim}
\end{figure*}

\subsection{Covariance matrix}\label{sec_cova}

The determination of the covariance matrix of the power spectrum is a key step towards obtaining constraints on cosmological 
parameters. Its determination represents a non-trivial task both from theoretical and numerical perspectives. From an analytical point of 
view, the covariance matrix can be decomposed into a Gaussian component, which depends on the measured power spectrum, and a non-Gaussian
contribution related to the bi- and tri-spectrum \citep{matarrese_bi,verde_tri,smith_cova}. On large scales, which are 
well described by linear perturbation theory, these non-Gaussian contributions vanish. 
On small scales however the non-linear evolution introduces coupling between Fourier modes generating signatures of non-Gaussianities.
Hence, to precisely determine the covariance matrix, it is necessary to have a model for these high-order statistics, together with 
redshift distortions, survey window function effects, beat-coupling effects \citep{hamilton_rimes,taka1} and correlations introduced 
by bin-averages \citep{meiksin}. From the numerical side, recent experiments with $N-$body simulations \citep[e.g.][]{taka1} have 
shown that large numbers of mock catalogues (or realizations) are required for a precise determination of the covariance matrix
as needed to constrain cosmological parameters.

We used our ensemble of 100 mock catalogues to obtain an estimate of the bin-averaged covariance matrix $\hat{C}(k_{i},k_{j})$
of the REFLEX II power spectrum  by 
\begin{equation}\label{variance_m}
\hat{C}(k_{i},k_{j})=\frac{1}{N_{\rm m}-1}\sum_{n=1}^{N_{\rm m}}\left( \hat{P}_{i}^{n}-\bar{P}_{i}\right)\left(\hat{P}_{j}^{n}-\bar{P}_{j}\right),
\end{equation}
where $\hat{P}_{i}^{n}=\hat{P}^{n}(k_{i})$ 
is the measured power spectrum in the $n$-th mock catalogue in the bin centered at $k_{i}$
and $\bar{P}_{i}$ is the mean power spectrum from the ensemble of mocks at the same bin. 
As an example, panel a) of Fig.~\ref{rc_2dim} shows examples of the correlation coefficients $r_{ij}$ defined from the covariance matrix via
$\hat{r}_{ij}=\hat{C}_{ij}/(\hat{C}_{ii}\hat{C}_{jj})^{1/2}$ for $L^{\rm min}_3$ (upper triangular part) and $L^{\rm min}_6$
(lower triangular part).
For comparison, panel c shows the correlation matrix inferred from the clustering of the illuminated halos in the L-BASICC II simulation
for the same luminosity cuts. The covariance matrix of the mocks contains important off-diagonal terms which
arise from the mode coupling induced by the window function. 
All the statistical analyzes performed in this work
are based on the covariance matrix defined by equation~(\ref{variance_m}).

\begin{figure} 
\includegraphics[width=8cm, angle=0]{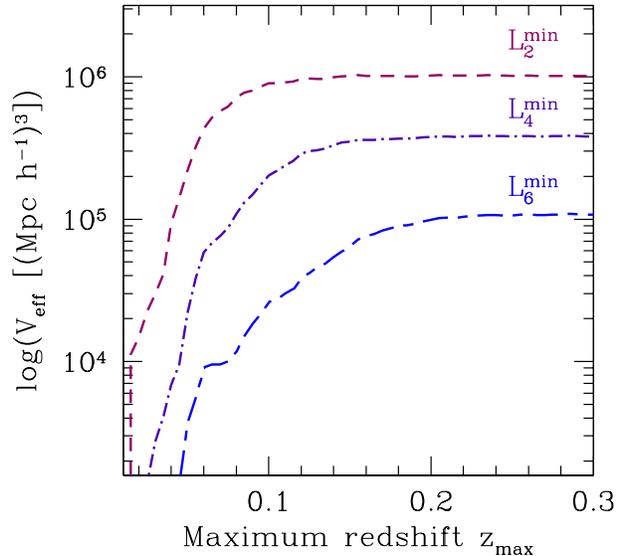}
\caption[]{Effective volume probed by the REFLEX II sample (see equation \ref{veff}) as a function of the maximum redshift for three different cuts 
in luminosity.}\label{eff_vol}
\end{figure}

\citet{FKP} derived an approximated expression for the variance of the spherically averaged power spectrum under the assumption that 
the Fourier modes are Gaussian-distributed. This is given by
\begin{equation}\label{sfkp}
\frac{\sigma^{2}(k)}{P(k)^{2}}=\frac{2}{V_{k}V_{\rm eff}(k)},
\end{equation}
where $V_{k}\approx 4\pi k^{2}\delta k/(2\pi)^{3}$ is the volume of a spherical shell of width $\delta k$ and $V_{\rm eff}(k)$
 is the effective (coherence) volume probed by the survey at a scale $k$, defined by \citet{teg_effv} as
\begin{equation}\label{veff}
V_{\rm eff}(k)=\int\left(\frac{\bar{n}(\textbf{r})P(k)}{1+P(k)\bar{n}(\textbf{r})}\right)^{2}\,{\rm d}^{3}r.
\end{equation}
Equation~(\ref{sfkp}) assumes that the power spectrum $P(k)$ is smooth on scales $\delta k$ and applies for wavenumbers
 $k_{i}\gg \delta k$. Fig.~\ref{eff_vol} shows the effective volume for different luminosity cuts as a function of the
 maximum redshift of the sample. 
As pointed out in Section \ref{sec_mocks}, the effective volume gained by including objects with redshifts $z\geq 0.22$
is very small and only leads to an increase of the shot-noise.

\begin{figure}
 \includegraphics[width=8.cm, angle=0]{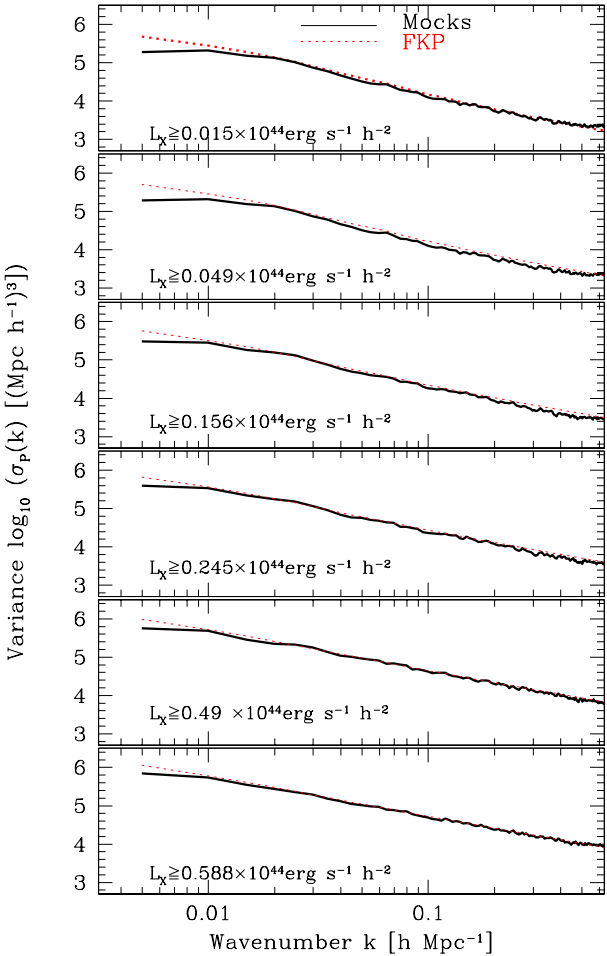}
 \caption[]{Bin-averaged standard deviation $\sigma_{p}(k)$: solid line represents the standard deviatation determined from the set of mock catalogues following equation 
(\ref{variance_m}). The dotted line shows the prediction form the equation (\ref{sfkp}).}\label{vpower}
\end{figure}

Fig.~\ref{vpower} shows a comparison of the theoretical variance computed using equation (\ref{sfkp}) with that derived from the
diagonal elements of the covariance matrix of the mocks (equation \ref{variance_m}) for the luminosity cuts defined in
Section~\ref{lfun}. The theoretical predictions were computed using a linear theory power spectrum with an amplitude
rescaled to match that of the REFLEX II measurements for the correspondent luminosity cut.
This comparison shows a very good agreement between the results obtained from the ensemble of mocks and the theoretical prediction for all 
values of $L^{\rm min}$. Equation~(\ref{sfkp}) thus provides a good estimate of the error bars in the measured power spectra.

\section{Understanding the REFLEX II power spectrum}\label{sec_under}

In this section we use our ensemble of mock catalogues to understand what can be expected from a measurement of the power
spectrum in the REFLEX II catalogue. Section \ref{lumi_bias} deals with a theoretical model for the luminosity dependence of the
bias. In Section \ref{sec_malm} we analyze the possible presence of a scale-dependent systematic bias induced by the flux-limited 
selection. Finally, Section \ref{sec_shape} describes a model for the full shape of the power spectrum including the 
distortions introduced by non-linear evolution.

\subsection{A theoretical prediction of luminosity bias}\label{lumi_bias}

Assuming that on the scales of interest the underlying halo-mass bias is scale-independent, the power spectrum of galaxy
clusters of a given luminosity $L_{X}$ and at a given redshift $z$ can be written as 
\begin{equation}
P_{\rm cl}(k,z;L_{X})=b^{2}(L_{X},z)P_{\rm mat}(k,z),
\end{equation}
where $P_{\rm mat}(k,z)$ is the matter power spectrum (we now drop the redshift dependence of these factors for simplicity).
The luminosity bias $b(L_{X})$ is given in terms of the halo mass function $n(M)$ and the underlying halo-mass bias 
$b(M)$ as \citep[e.g.][]{hmodel_best} 
\begin{equation}\label{ebll}
b(L_{X})=\frac{\int \,  n(M)b(M)p(L_{X}|M)\, {\rm d} M}{\int \, n(M)p(L_{X}|M)\, {\rm d} M}.
\end{equation}
Here $p(L_{X}|M)$ represents the conditional probability distribution of assigning a X-ray luminosity $L_{X}$ to a dark matter halo of 
mass $M$. 
As described in Section \ref{sec_mocks}, for our mock catalogues $p(L_{X}|M)$ is given by a log-normal distribution with a mean given
by equation (\ref{eq:teo_mass_lum}) and a scatter including an intrinsic dispersion $\sigma_{\rm ln L}=0.26$ and a 
$20$ per cent luminosity error added in quadrature.
In equation (\ref{ebll}), the numerator is the theoretical prediction for the X-ray luminosity function
\begin{equation}\label{eq:flum_theo}
 \Phi(L_{X})=\int\, n(M)p(L_{X}|M)\,{\rm d} M,
\end{equation}
 given by the halo mass function and the mass-luminosity relation. 

\begin{figure} 
\includegraphics[width=8cm,angle=0]{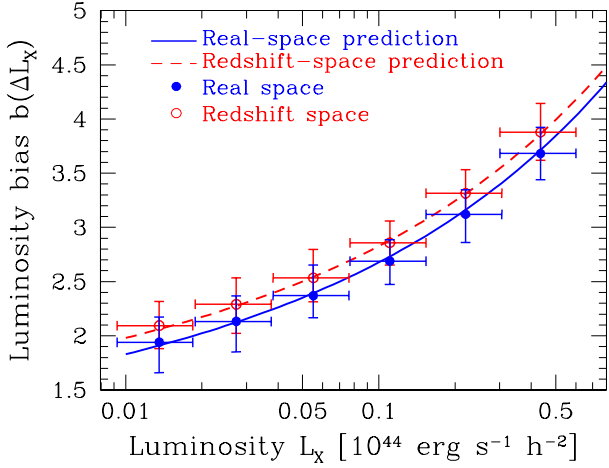}
\caption[]{\footnotesize{Measured luminosity bias $\hat{b}(\Delta L_{X})$ from the L-BASICC II simulation in real space 
(filled points) and redshift space (open points). Solid and dashed lines represent the prediction from equations (\ref{ebll}) and (\ref{beef}) respectively. }}\label{lb_lbasicc}
\end{figure}

In order to test the predictions of equation (\ref{ebll}), we used the $50$ realizations of the L-BASICC II to measure the 
absolute luminosity bias. We assigned luminosities to the dark matter haloes using the calibrated mass-luminosity relation described 
in Section \ref{sec_mocks}. We then measured the halo power spectrum $\hat{P}_{\rm cl}(k;\Delta L_{X})$, in real-space,
by splitting the sample in luminosity bins of width $\Delta L_{X}$. We used the measurements of
the real-space dark matter power spectrum of the same simulations to determine the ratios 
\begin{equation}
\hat{b}(\Delta L_{X})\equiv \sqrt{\frac{\hat{P}_{\rm cl}(k,\Delta L_{X})}{\hat{P}_{\rm mat}(k)}}.
\label{bias_hat}
\end{equation}
For this we used Fourier modes in the range $0.01\leq k /(h\, {\rm Mpc}^{-1})\leq 0.1$, where this ratio is compatible with a
scale-independent bias factor. The obtained measurements are shown by the filled circles in Fig.~\ref{lb_lbasicc}.

The measured bias factors can be estimated by integrating equation~(\ref{ebll}) within the given luminosity bin 
\begin{equation}\label{bxl}
b(\Delta L_{X})=\frac{\int_{\Delta L_{X}}\Phi(L_{X})b(L_{X})\,{\rm d} L_{X}}{\int_{\Delta L_{X}}\Phi(L_{X})\,{\rm d} L_{X}},
\end{equation}
where $\Phi(L_{X})$ is given by equation~(\ref{eq:flum_theo}).
This prediction is shown by the solid line in Fig.~\ref{lb_lbasicc}.
For this we used the mass function from \citet{jenkins} and the halo bias from \citet{tinker_b}, which provide an excellent description
of these quantities in the L-BASICC II simulations. There is an excellent agreement between the theoretical $b(\Delta L_{X})$
and the measurements obtained from the simulations. This agreement depends on the accuracy with which the $M-L_{X}$ relation is known.

Under the assumptions of linear evolution and the distant observer approximation, 
the cluster power spectrum in redshift-space is a boosted version of its real-space counterpart,
$P_{\rm cl}^{s}(k;\Delta L_{X})=S(\Delta L_{X})P_{\rm cl}(k;\Delta L_{X})$ \citep{kaiser}, where 
\begin{equation}\label{beef}
S(\Delta L_{X})=1+\frac{2}{3}\frac{f}{b(\Delta L_{X})}+\frac{1}{5}\left(\frac{f}{b(\Delta L_{X})}\right)^{2},
\end{equation}
where $f\equiv {\rm d}\ln D(a)/{\rm d}\ln a$ is the growth index \citep[e.g.][]{wang} and $D(a)$ is the growth factor \citep[e.g.][]{peebook}. The open circles in Fig.~\ref{lb_lbasicc} show the bias factors $b^{\rm s}(\Delta L_{X})$ estimated from the L-BASICC II
simulations as in equation~(\ref{bias_hat}) using the redshift-space halo power spectrum  $\hat{P}^{\rm s}_{\rm h}(k,\Delta L_{X})$.
The dashed line in this figure corresponds to the theoretical prediction for this quantity, given by $b(\Delta L_{X})^2S(\Delta L_{X})$.
The agreement between this prediction and the results from the $N$-body simulations show the validity of this simple
treatment of the redshift space-distortions.

\begin{figure}
  \includegraphics[width=8cm, angle=0]{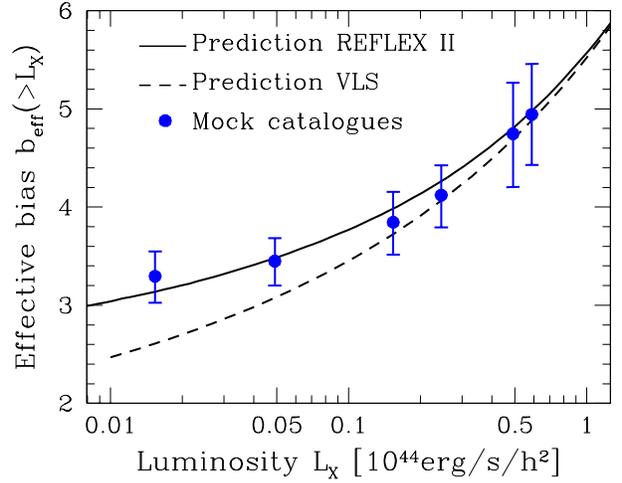}
  \caption[]{Effective luminosity bias measured from the REFLEX II mock catalogues (filled circles with error bars). The solid line
shows the prediction from equation (\ref{beff3}) with the fiducial values $\sigma_{\ln L}=0.26$ and constant luminosity error of
$20$ per cent. The dashed line is the prediction for a VLS.}\label{eff_biasm}
\end{figure}
The results of equations (\ref{bxl}) and (\ref{beef}) can also be used to estimate the effective bias of a 
power spectrum measurement in the REFLEX II catalogue for a given luminosity cut $L^{\rm min}$. In order to achieve this, we first write the cluster power spectrum in redshift-space at a given redshift $z$ in terms of the linear dark-matter power spectrum at $z=0$ as 
\begin{equation}
P^{\rm s}_{\rm cl}(k,z;>L^{\rm min})=b^{s}(>L^{\rm min},z)^{2}P_{\rm mat}(k,z=0),
\end{equation}
where 
\begin{equation}
 b^{\rm s}(z,>L^{\rm min})^{2}=b(>L^{\rm min},z)^{2}S(z,>L^{\rm min})D^{2}(z).
\end{equation}
In this expression we take into account that the minimum luminosity included in the sample
varies with $z$ following the REFLEX II selection function. Accordingly, we determine the bias $b(z,>L^{\rm min})$ following equation~(\ref{bxl}) with the red-shift dependence given by the REFLEX II sensitivity map:
\begin{equation}\label{beff2}
b(z,>L^{\rm min})=\frac{1}{\bar{n}(z,>L^{\rm min})}\int_{\hat{L}(z)}^{\infty}  \Phi(L)b(L)\,{\rm d} L\,, 
\end{equation}
where the mean number density $\bar{n}(z,>L^{\rm min})$ and the lower integration limit $\hat{L}(z)$ are given by equations (\ref{nbar})
and (\ref{lhat}) respectively. Note that $\hat{L}(z)$ should not only depend on the redshift but also on the angular position, according to the REFLEX II
sensitivity map. However, in order to give an estimate of the effective bias at a fixed redshift we made an average of the values of
$L^{\rm min}(\textbf{r})$ of all the $N_{\rm pix}$ pixels within the REFLEX II mask. Individual pixels displayed small differences
compared to the average in the final result. 

The effective bias of the full sample will then be given by the average of the bias factors 
of equation (\ref{beff2}) over the observed volume as \citep[e.g.][]{suto,moscardini}
\begin{equation}\label{beff3}
b_{\rm eff}(>L^{\rm min})^{2}=\frac{\int_{z}[\bar{n}(z,>L^{\rm min})b^{\rm s}(>L^{\rm min},z)]^{2}\,\frac{{\rm d}V}{{\rm d}z}\,{\rm d}z }
{ \int_{z}\, \bar{n}(z,>L^{\rm min})^{2}\frac{{\rm d}V}{{\rm d}z}\,{\rm d}z},
\end{equation}
where the integrals are evaluated in the redshift interval of the sample with a volume element ${\rm d}V/{\rm d}z=r(z)^{2}/H(z)$
according to our fiducial cosmology. This therefore allows us to write a prediction for the observed power spectrum from a sub-sample characterized by the luminosity cut $L^{\rm min}$ as 
\begin{equation}
P_{\rm cl}(k;>L^{\rm min})=b_{\rm eff}(>L^{\rm min})^{2}P_{\rm mat}(k,z=0).
\end{equation}
The solid line in Fig.~\ref{eff_biasm} shows the prediction for the effective bias of the REFLEX II catalogue as a function of the minimum luminosity $L^{\rm min}$ computed using equation~(\ref{beff3}) (setting $D(z)=1$ since, by construction, the mock catalogues assume no redshift evolution). For comparison, the dashed line in Fig.~\ref{eff_biasm} represents the equivalent prediction for a volume limited sample. Equation~(\ref{beff3}) gives an excellent description of the direct measurements
obtained from the mock catalogues (shown by the filled circles). Then, this model provides a means to extract the important cosmological information contained in the amplitude of the measured REFLEX II power spectrum.

\subsection{Systematics: distortions induced in a flux limited sample}\label{sec_malm}
\begin{figure}
  \includegraphics[width=8cm,angle=0]{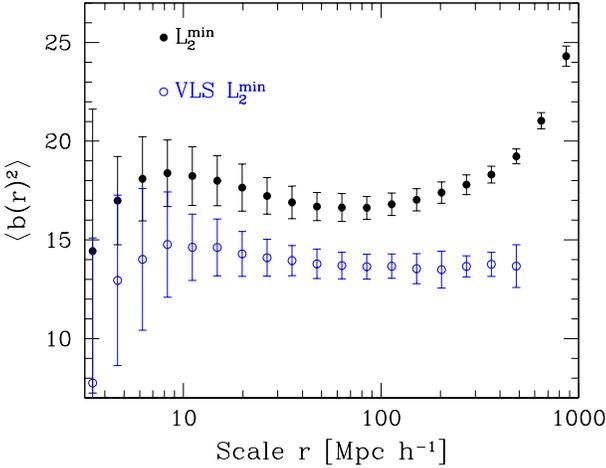}
  \caption[]{Mean squared luminosity bias for pairs separated by a scale $r$ in the REFLEX II mock catalogues, for the sub-sample $L^{\rm min}_{2}$ (filled circles) and its corresponding VLS (open circles).}\label{b_scale}
\end{figure}

Due to the flux-limited nature of the REFLEX II catalogue, large scales might be probed predominantly by the most luminous clusters,
with a higher clustering amplitude. This would artificially increase the measured power spectrum on large scales, introducing a 
scale-dependent distortion with respect to the volume limited case. \citet{reflex1} analyzed this problem in detail for the
REFLEX sample and concluded that no significant effect can be detected for scales $r < 150\,{\rm Mpc}\,h^{-1}$.
In this section we perform a similar analysis on the REFLEX II sample.  
As the REFLEX II sample spans a wider range of luminosities, and covers a larger volume than that used by \citet{reflex1}, it is
necessary to test whether this systematic effect can affect our measurements.

Fig.~\ref{b_scale} shows the number of pairs with separation $r$ weighted by the
individual biasing factors of the pair members (computed using equation~(\ref{ebll})) determined from the mock catalogues,
yielding the average squared bias factors
\begin{equation}\label{mark}
\langle b^{2}(r)\rangle = \frac{1}{n(r)}\sum_{i,j}b(L_{i})b(L_{j}), 
\end{equation}
where the sum is done over pairs separated by scales in the range $r-\frac{1}{2}\Delta<|r_{i}-r_{j}|<r+\frac{1}{2}\Delta$ and $n(r)$ is 
the number of clusters in the same interval. The closed circles show the results obtained from the sample with 
minimum luminosity $L_2^{\rm min}$ (see Table~\ref{tableVLS}), while the open circles correspond to the equivalent measurement 
from the VLS determined by the same minimum luminosity. The error bars are drawn from the standard deviation of the ensemble.
Two prominent features can be observed from Fig.~\ref{b_scale}. On one hand, there is an increase in the mean bias on scales $r\approx 10 {\rm Mpc}\, h^{-1}$. This is understood as to show that pairs of clusters separated by these scales have at least one cluster with luminosity bias higher than the mean value of the sample. This is therefore a consequence of gravitational clustering. On scales smaller than  $r \sim 9 {\rm Mpc}\,h^{-1}$ the mean bias decreases as a consequence of the halo-exclusion \citep[e.g.][]{porciani}. This feature is more evident in the sub-sample $L^{\rm min}_{2}$ than in its corresponding volume-limited counterpart, due basically to the low volume sampled by the latter. On the other hand, it can be clearly seen that, on scales $r \gtrsim 150\,{\rm Mpc}\,h^{-1}$, 
a systematic increase of the average squared bias factor exists in the flux-limited sample compared with the volume-limited case. This is a direct consequence of the flux-limited nature of the survey. Naively, the scale where this flux-selection distortions is relevant would correspond to 
$k =2\pi/r = 0.04\,h\,{\rm Mpc}^{-1}$, suggesting that for wavenumbers larger than this limit
no scale-dependent distortion affects the measurements from the REFLEX II catalogue. Comparing with other sub-samples, we observe that this distortion is damped in the high luminosity cuts, which almost behave like volume-limited samples (see Fig.~\ref{lz_vls}).

\begin{figure}
  \includegraphics[width=8cm, angle=0]{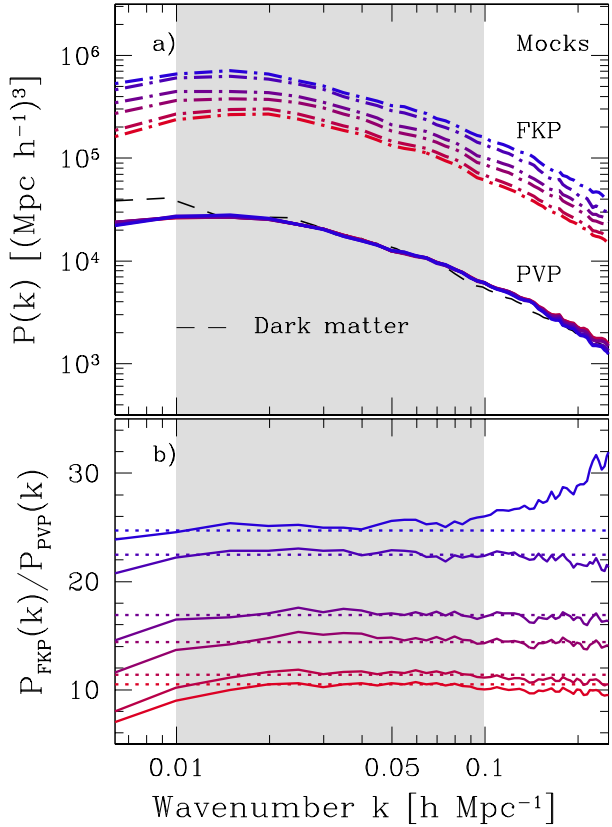}
  \caption[]{a) Comparison of the resulting mean power spectrum from the mock sample using the PVP estimator (solid lines) and the
FKP result (dot-dashed lines) for our six cuts in luminosity defined in Table~\ref{tableVLS}. The dashed line represents the measurement of the dark matter power
spectrum of the L-BASICC II simulations. The shaded region shows the scales where the absolute
bias is fitted ($0.01\leq k\, (h\,{\rm Mpc}^{-1})\leq 0.1$) b) Ratio of FKP power spectra to the PVP measurements for the six luminosity cuts.}\label{reflex2_power_pvp}
\end{figure}

To analyze this issue in more detail we used the FKP estimator as implemented by \citealt{pvp} (hereafter PVP).
This is also a minimal variance weighting estimator which takes into account the absolute luminosity bias (or the relative bias to some luminosity $\tilde{L}$) to obtain an estimate of the power spectrum of the underlying matter (or correspondingly, the power spectrum of objects with luminosity $\tilde{L}$), free of the distortions induced by the flux-limited selection of the sample. Recently, \citet{cai} generalized this estimator by introducing a mass-dependent weighting scheme that minimizes the stochasticity between the cluster (or galaxy) field and the underlying matter distribution. For the goal of this section it is enough to use the PVP estimator.

Panel a) in Fig.~\ref{reflex2_power_pvp} shows a comparison of the mean power spectra obtained by the FKP (dot-dashed lines)
and PVP (solid lines) algorithms for the different luminosity cuts defined in Table~\ref{tableVLS} in our ensemble of mock catalogues. 
In the PVP method, we weighted each object by the inverse of its luminosity bias, computed using a fit to the 
redshift-space results shown in Fig.~\ref{lb_lbasicc}. The shaded area in Fig.~\ref{reflex2_power_pvp} represents the range of scales
used to measure these bias factors ($0.01\leq k\,(h {\rm Mpc}^{-1})\leq 0.1$). This produces a power spectrum 
normalized as that of the dark matter distribution (shown by the dashed line in Fig.~\ref{reflex2_power_pvp}). 
Panel b) shows the ratios between the mean power spectra obtained using the FKP and PVP estimators for the each luminosity cut.
In the absence of a scale-dependent distortion, these ratios should correspond to the bias factors $b^2_{\rm eff}(>L_{\rm min})$, 
shown by the dotted lines. These ratios show no clear signature of a scale-dependent distortion for $k)\leq 0.02\,h\,{\rm Mpc}^{-1}$. 
Contrary to what might be expected from this systematic effect, Fig.~\ref{reflex2_power_pvp} shows a weak indication of a 
decrease in power on large scales for the lower luminosity cuts, which is smaller than the standard deviation of the measurements.
From this analysis we conclude that no significant distortion is introduced in the shape of the power spectra estimated with 
the FKP method. Note however that this statement applies exclusively to the REFLEX II catalogue, as such an effect has been detected
in galaxy surveys \citep[e.g.][]{teg_sdss,percival_sdss}.
\begin{figure}
  \includegraphics[width=8cm, angle=0]{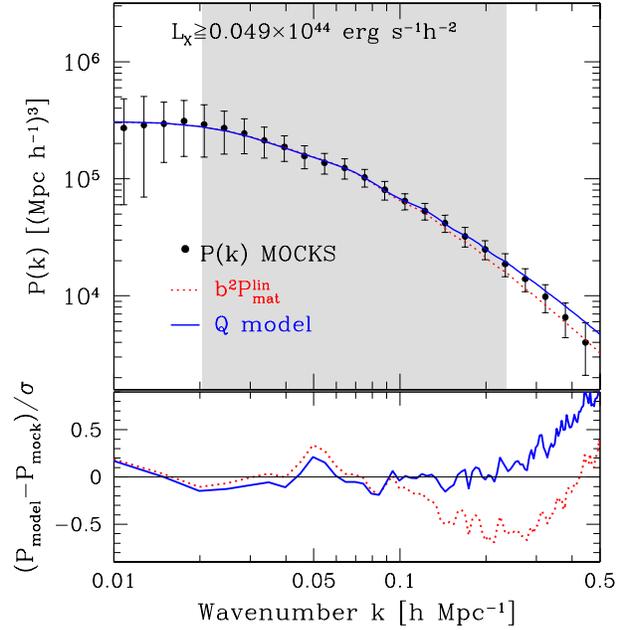}
  \caption[]{$Q$-model description of the mock power spectrum. The shaded region shows the scale where the $Q$-model was used to fit
the mean power spectrum from the mock catalogues. The bottom panel show the ratio of the difference between the $Q$-model and the measurements to
the standard deviation from the mocks.}\label{reflex2_power_qmodel}
\end{figure}
Given the lack of systematic distortions in the FKP measurements, we chose to use the FKP estimator to analyze the 
data from the REFLEX II catalogue. Panel b) of Fig.~\ref{rc_2dim} shows the correlation matrix inferred from the ensemble of mock catalogues of the power spectra for two luminosity cuts obtained using the PVP estimator. A comparison with panel a) shows that the PVP estimator induces slightly higher correlations between Fourier modes compared to the FKP case.

\subsection{Modeling the shape of $P(k)$}\label{sec_shape}

In this section we use our ensemble of mock catalogues to test a model of the shape of the REFLEX II power spectrum. 
We focus on the clusters with luminosities greater than $L^{\rm min}_{2}$. The filled circles in Fig.~\ref{reflex2_power_qmodel}
show the mean redshift-space power spectrum of the mocks for this luminosity cut with error-bars determined from the standard deviation
of the ensemble. It can be clearly seen that this measurement exhibits an excess of power at small scales with respect to the
predictions from linear perturbation theory, shown by the dashed line. This is due to the combined effect of non-linear
evolution and redshift-space distortions. 

In recent years the distortions in the shape of the power spectrum produced by these effects have been intensively studied 
using large $N$-body simulations and recent advances in perturbation theory \citep[e.g.][]{sss,rpt,angulo,ariel1,montesano}.
These analyses have produced accurate descriptions of these distortions to the level of accuracy demanded by forthcoming surveys,
which will probe volumes much larger than present day catalogues. Due to the moderate volume probed by the REFLEX II catalogue (in comparison to the volume probed by current galaxy redshift surveys), percent-level accuracies in the treatment of these effects are not required. 

\begin{figure*}
  \includegraphics[width=15cm, angle=0]{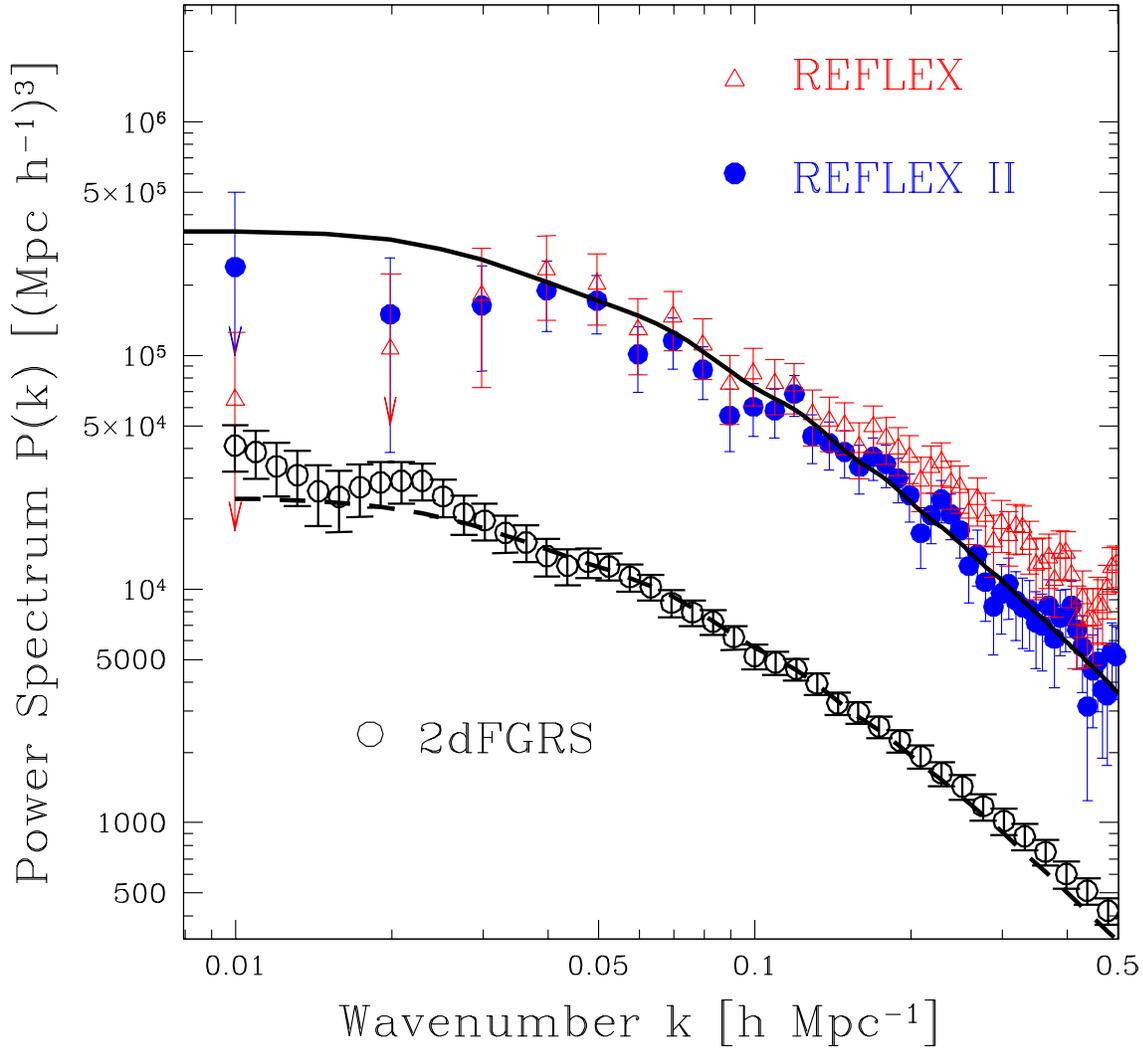}
  \caption[]{REFLEX II power spectrum (filled circles with error bars) for clusters with luminosities $L_{X}>L^{\rm min}_{1}$. The REFLEX power spectrum is shown by the open triangles. The error bars for these two measurements are taken from equation~(\ref{sfkp}). For comparison we also show the
measured power spectrum from the $2$dfGRS taken from \cite{cole2df} (open circles). The solid and dashed line represent the $\Lambda$CDM power spectrum
convolved with the REFLEX II and the $2$dfGRS window function respectively, and adjusted to match the corresponding spectra. Error-bars exceeding the range of the plot are represented by arrows.}\label{comparison}
\end{figure*}

\begin{figure*}
  \includegraphics[width=17cm, angle=0]{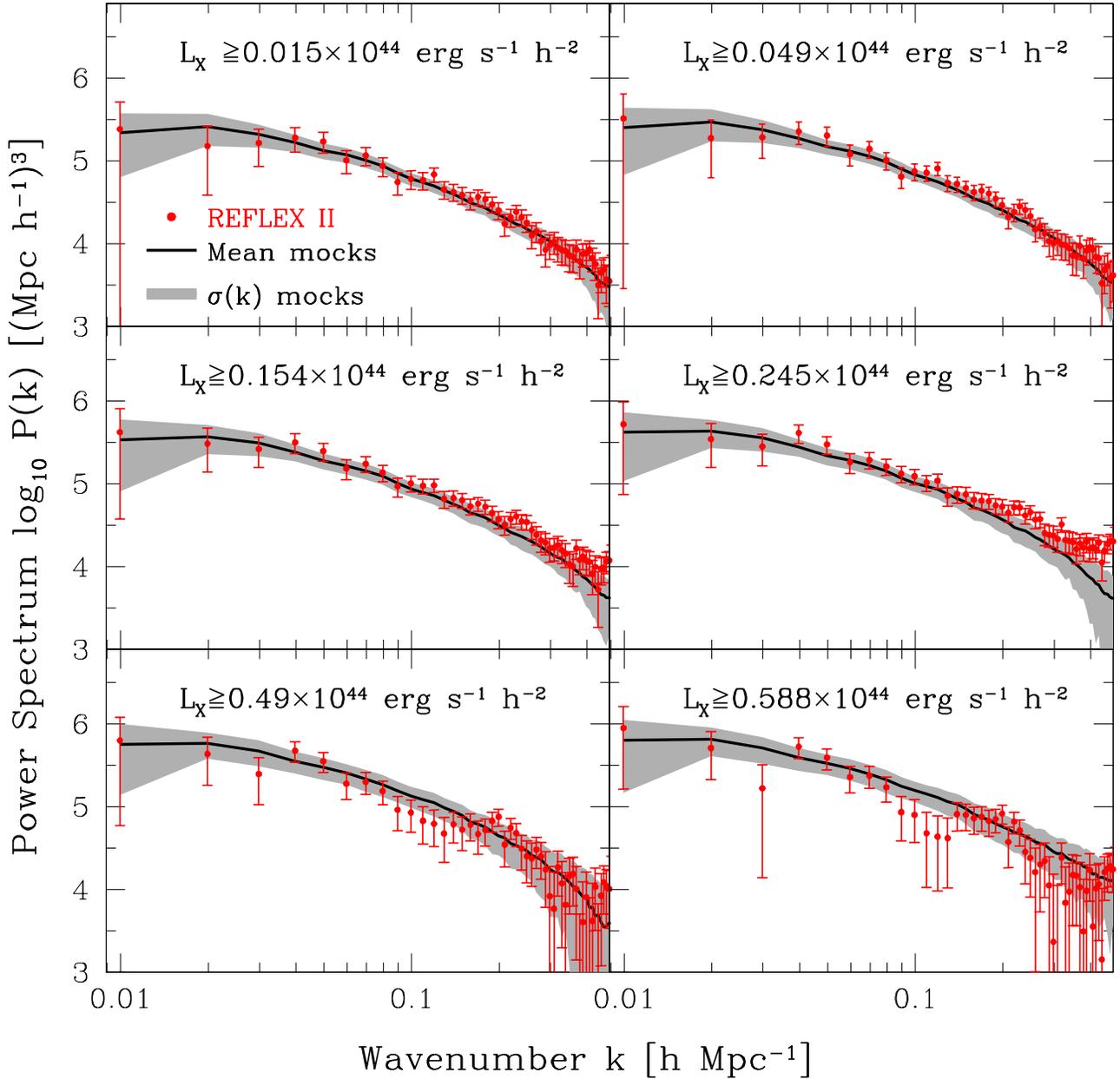}
  \caption[]{Measured REFLEX II power spectrum for the different sub-samples defined in Table~\ref{tableVLS}. FFTw is carried out in a box of side $L_{\rm box}=1263.8\,{\rm Mpc}\,h^{-1}$
and $P_{\rm est}=2\times 10^{4}({\rm Mpc}\,h^{-1})^{3}$. The fundamental mode is $\delta k=2\pi/L_{\rm box}=0.0049 \,h\,{\rm Mpc}^{-1}$ and the Nyquist frequency is $k_{\rm N}=1.27\,h\,{\rm Mpc}^{-1}$. Points represent the
REFLEX II measurements with error bars drawn from equation~(\ref{sfkp}). The shaded region represents the $1\sigma$ standard deviation determined from the
mocks catalogues. The solid line represents the mean mock power spectrum.}\label{power_mock}
\end{figure*}

We now test whether the $Q$-model of \citet{cole2df} \citep[modified as in][]{ariel1} can provide a good
description of the non-linearities observed in the power spectra of our mock catalogues.  In this model the shape of the cluster power spectrum is given by
\begin{equation}\label{eq:qmodel}
P_{\rm cl}(k,>L)=b_{\rm eff}(>L)^{2}\left( \frac{1+Qk ^{2}}{1+Ak+Bk^{2}}\right) P^{\rm lin}_{\rm mat}(k),
\end{equation}
where $P^{\rm lin}_{\rm mat}(k)$ is the linear theory matter power spectrum. 

Although this model was originally developed and calibrated to
to describe the power spectrum of the 2dFRGS, its application
has been extended to the analysis of other samples \citep[e.g.][]{teg_sdss,padma}.
In particular, \citet{sanchez_cole} showed that this model can give a
good description of the
clustering of the LRG sample from SDSS even though it was not
specifically designed to do so.
At the same time this model does not give a good description of the
shape of $P(k)$ for the main galaxy sample in SDSS.
The results from the application of the $Q$-model to $N$-body simulations
show that it can correctly describe the clustering of dark matter halos above a given mass threshold \citep{teg_sdss}.

We follow \citet{cole2df} and fix the value of $A=1.4$ as obtained from the analysis of $N$-body simulations, while $Q$ and $B$
are left as free parameters whose values will depend on the limiting luminosity of the sample. 
We assumed all the cosmological parameters to be known and fitted for $Q$ and $B$ marginalyzing analytically
over the amplitude \citep[as described in][]{lewis}. From this analysis we obtain the values $Q=24.9\pm 1.1$ and
$B=12.0\pm 2.1$, corresponding to the sub-sample defined by $L^{\rm min}_{2}$. The best fit model obtained this way is shown by the solid line in Fig.~\ref{reflex2_power_qmodel}. It can be clearly seen that the model of equation~(\ref{eq:qmodel})
gives an accurate description of the shape of the 
mean power spectrum from our ensemble of mock catalogues. This can be also seen in panel b) of the same figure, where we show the ratio between the difference of the mean mock power spectrum and the best fit-model to the variance from the ensemble. 
The parameters $B$ and $Q$ fitting the power spectrum of the sub-sample $L^{\rm min}_{2}$ follow a degeneracy that can be described approximately by $B(Q)=0.805Q-8.15$. This degeneracy is maintained if the amplitude of the model is fixed according
to equation~(\ref{beff3}). We can thus use this degeneracy to reduce the number of degrees of freedom when constraining cosmological parameters using the measured power spectrum. The best fit value of $Q$ increases with the limiting luminosity
of the sample, varying from $Q=20.7\pm 0.9$ for $L^{\rm min}_{1}$ to $Q=44.9\pm 2.3$ for $L^{\rm min}_{6}$. 
The general trend in the degeneracy $B(Q)$ is maintained for different luminosity cuts. In Section~\ref{sec:shape_r}
we compare the predictions of this model with the measurement of the REFLEX II power spectrum. We have explictly tested that, by adopting a different value of A the best fit value for the parameter Q, and the
degeneracy between Q and B, are slightly changed, while providing equally good fits to the data.

\section{Analysis of the REFLEX II power spectrum}\label{sec_q}

\subsection{Measurements}\label{sec:measure}

The measured power spectrum for the REFLEX II sample with limiting luminosity $L^{\rm min}_{1}$ is shown by the 
filled points in Fig.~\ref{comparison} with
error bars drawn from the FKP method (see equation~\ref{sfkp}). The solid line represents a $\Lambda$CDM linear power spectrum,
convolved with the window function of the survey. This theoretical prediction was computed using the fitting formulae of \citet{hu}, with amplitude rescaled to match that of the REFLEX II measurement. This simple exercise shows that the shape
of the REFLEX II power spectrum is consistent with the predictions of the $\Lambda$CDM cosmological model. 

Fig.~\ref{comparison} also shows a new estimation of the power spectrum of the original REFLEX sample (open triangles).
The REFLEX power spectrum has a higher amplitude, as expected from the higher flux-limit of this sample, and its shape
is in good agreement with that of the REFLEX II measurement.
The larger volume probed by the new catalogue reduces the impact of cosmic variance on large scales, 
where the REFLEX II power spectrum exhibits a higher amplitude than the measurement in the original REFLEX sample.
 
Fig.~\ref{comparison} also shows the galaxy power spectrum measured from the $2$dFGRS \citep{cole2df}.
The dashed line represents the same $\Lambda$CDM power spectrum described above convolved with the 2dFGRS window function.
This shows that, once their respective window functions have been taken into account, the large-scale ($k < 0.1\, h\,{\rm Mpc}^{-1}$)
shape of the power spectra inferred from the REFLEX II and the 2dFGRS are in good agreement and can be described with the same
cosmological model. At smaller scales, redshift-space distortions and non-linear evolution produce deviations in the shapes
of these power spectra.

Fig.~\ref{power_mock} shows a comparison of the measured REFLEX II power spectra (points with error bars) for 
the six cuts in luminosity described in Section \ref{lfun}, and the corresponding mean power spectra from the mock catalogues
(solid lines), with their corresponding $1\sigma$ {standard deviation} (shaded regions). The error bars of the REFLEX II power spectrum
correspond to the theoretical prediction of the FKP method (see section \ref{sec_cova}). We observe that the spectra measured in the
mocks are compatible within $1\sigma$ with the REFLEX II clustering up to $k\approx 0.3\,h\,{\rm Mpc}^{-1}$ for all luminosity cuts.
Notice that the mocks were only calibrated to follow the X-ray luminosity function of the REFLEX II sample. 
This agreement allows us to use the covariance matrices inferred from the ensemble of mock catalogues when analyzing the 
REFLEX II measurements.

On small scales the power spectra inferred from the mock catalogues are affected by the halo exclusion effect.
Dark matter halos in the simulations have been counted as separate entities when they did not overlap with their radii
of $r_{\rm FoF}$. On the other hand, in the REFLEX II catalogue clusters
have been treated as distinct if their X-ray emission does not significantly overlap.
Due to the short exposures in the RASS, the outer boundary of the X-ray emission (in two dimensional images, which is
significantly smaller than the aperture radius determined from one-dimensional profiles) is smaller than the radii of $r_{500}$.
This produces differences between the REFLEX II power spectra and the results from the mock catalogues on scales 
$k > 0.2\,h\,{\rm Mpc}^{-1}$.

Regarding the redshift incompleteness of the REFLEX II catalogue (around $10$ per cent), we verified that our
results are not substantially modified when the power spectrum is measured after randomly subtracting up to $20$ per cent
of the total number of clusters.

 \begin{figure}
 \includegraphics[width=8cm, angle=0]{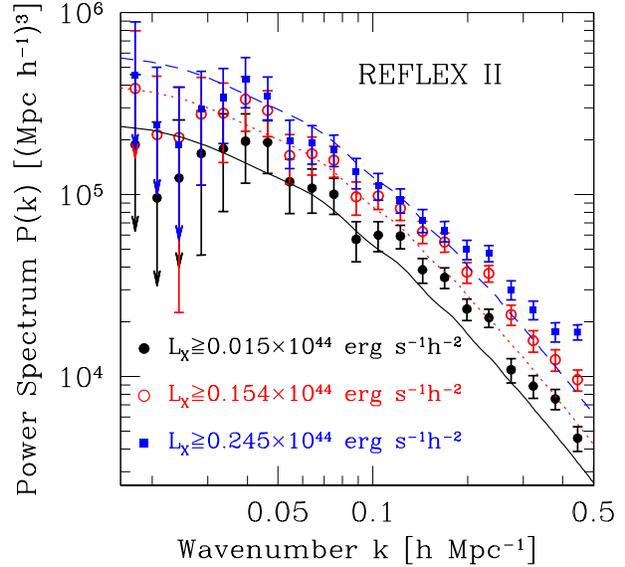}
 \caption[]{REFLEX II power spectrum for three different minimum luminosities. Lines represent a linear-perturbation theory power spectrum with our fiducial cosmology and amplitude given by equation~(\ref{beff3}). Error bars correspond to the {standard deviation} drawn from the mock catalogues. Arrows denote the error bars exceeding the range of the plot.}\label{Lpower}
\end{figure}

\subsection{Amplitude of the REFLEX II $P(k)$}\label{sec_amp}
Fig.~\ref{Lpower} shows the measurements of the REFLEX II power spectra for three of the sub-samples defined in Table~\ref{tableVLS}. The increase in the amplitude with increasing minimum luminosity can be clearly seen, showing the
signature of luminosity bias. 

\begin{figure} 
  \includegraphics[width=8.0cm,angle=0]{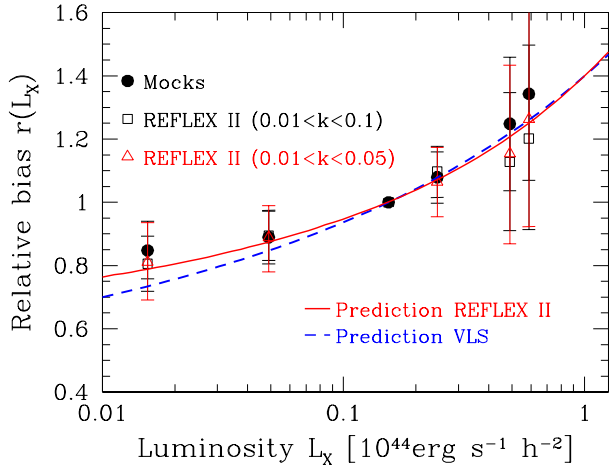}
\caption[]{Relative luminosity bias measured from the REFLEX II data in the range $0.01<k\,(h\,/{\rm Mpc})<0.1$ (open squares) and $0.01<k\,(h\,/{\rm Mpc})<0.05$ (open triangles) and the mock catalogues (filled points). The reference luminosity is $\tilde{L}=L^{\rm min}_{3}$. The dashed line is the prediction from equation (\ref{rbm}) for a VLS, while the solid line is the prediction from equation (\ref{beff3}) used in equation~(\ref{rbm}).}\label{bl1}
\end{figure}

In Section~\ref{lumi_bias} we showed that the measurements of the effective bias of the REFLEX II mock
catalogues are well described by the predictions of equation~(\ref{beff3}). In this section we confront this prediction with 
the power spectra measured from the REFLEX II sample. In order
 to avoid using the underlying dark matter power spectrum, we test equation~(\ref{beff3}) by means of the a relative luminosity bias $r(L_{X})$, defined as the ratio between the power spectrum of the sub-sample defined by a minimum luminosity $L_{X}$ to that of clusters with luminosities greater than a reference value $\tilde{L}_{X}$:  
\begin{equation}\label{rbm}
r(L_{X})\equiv \frac{b_{\rm eff}(>L_{X})}{b_{\rm eff}(>\tilde{L}_{X})}.
\end{equation}
The results are shown in Fig.~\ref{bl1}. The open squares show the measurements from the REFLEX II data and the filled points
correspond to the respective measurement from the mock catalogues. The solid line shows the prediction from equation~(\ref{beff3}), while the dashed line is the prediction of the effective bias for a VLS. The prediction from equation (25) provides a good description of the
REFLEX II measurements in the low luminosity cuts. For the last two luminosity cuts the
agreement is not as good, although the
measured bias factors are consistent with the theoretical prediction
within the errors.
As can be seen in Fig.~\ref{power_mock}, the spectra measured from these luminosity
cuts show a low amplitude at scales $k\sim 0.1 h/{\rm Mpc}$. This dip in the power spectra,
which is consistent with cosmic variance,
explains why the correspondent bias measurements in Figure 16 lie
below the theoretical prediction.
The agreement is improved when the bias factors are measured in the
range $0.01 < k (h /{\rm Mpc}) < 0.05$, as is shown by
the open triangles in Figure 16. These results confirm the validity of
equation \ref{beff3} to
model the amplitude of the REFLEX II power spectrum.

Note, finally, that as the theoretical predictions for the bias
are based on the mass-X-ray luminosity
relation calibrated from the mock catalogues, the good agreement found
with the bias factors inferred from
the data suggests that for the luminosity range we have explored, the
calibrated scaling relation provides
also a good description of the real underlying mass-X ray luminosity relation.

\subsection{Shape of the REFLEX II $P(k)$}\label{sec:shape_r}
\begin{figure}
  \includegraphics[width=8cm, angle=0]{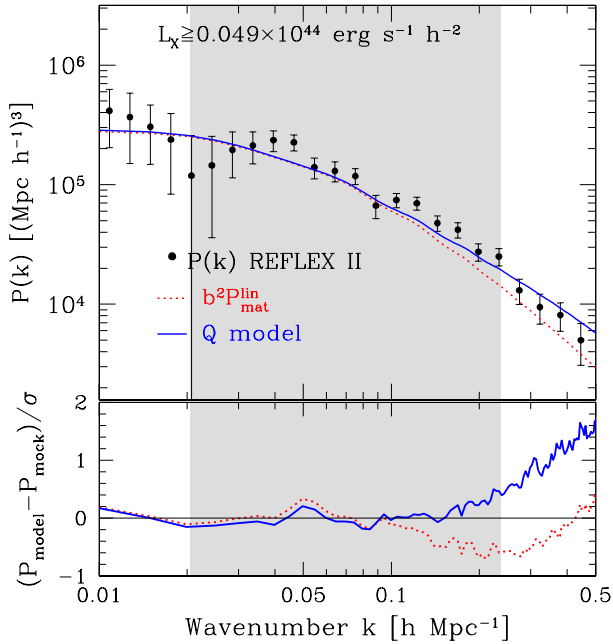}
  \caption[]{Best-fitting  $Q$-model for the REFLEX II mock power spectrum (points with error bars). See Fig.~\ref{reflex2_power_qmodel} for description.}\label{reflex2__qmodel}
\end{figure}

We used the $Q$-model to analyze the shape of the REFLEX II power spectrum for the sub-sample defined by 
$L^{\rm min}_{2}$, following the same procedure as in Section~\ref{sec_shape}. The best-fit of the $Q$-model is shown by the solid line in the upper panel of Fig.~\ref{reflex2__qmodel}. For this measurement we find $Q=24.7 \pm 1.5$ and $B=8.6\pm1.1$, with a degeneracy described by $B(Q)=0.72Q-9.25$. This degeneracy is maintained when fixing the amplitude of the model 
power spectrum according to equation~(\ref{beff3}). As in Fig.~\ref{reflex2_power_qmodel}, the bottom panel shows the ratio
of the difference between the model and the measurements to the standard deviation determined from the mock catalogues. This shows that
a model including a correction for non-linearities provided by equation~(\ref{eq:qmodel}) gives a better description of the shape of the REFLEX II power spectrum than the predictions from linear perturbation theory.

\section{Conclusions}\label{sec_conc}

In this paper we presented the measurement and analysis of the power spectrum from the new REFLEX II catalogue, which 
is an extension of the original REFLEX sample \citep{hb_catalogo} to a lower limiting flux 
($1.8 \times 10^{-12}\, {\rm erg}\,{\rm s}^{-1}\,{\rm cm}^{-2}$).
The new sample contains $911$ X-ray detected galaxy clusters of which $860$  have measured redshifts in the range
$0\leq z\lesssim 0.6$ and X-ray luminosities in the range  $4.9\times 10^{40}\leq L_{X}/({\rm erg}\,\,{\rm s}^{-1}h^{-2})\leq 1.96\times 10^{45}$. The total flux and X-ray luminosities are estimated using the up-to-date scaling
relations based on the REXCESS Survey \citep{bohringer_07,pratt}.

The new sample allowed us to perform a detailed study of the full shape and amplitude of the power spectrum of 
X-ray detected galaxy clusters. We complemented this analysis by using a set of 100 independent mock catalogues
constructed to match the selection function of the REFLEX II survey. 
The clustering properties of these mock catalogues are in good agreement with those measured in the REFLEX II sample.
Thus, this ensemble provides a reliable tool to test the statistical methods applied to the data. In particular, we used
the mock catalogues to test a model for the luminosity dependence of bias, to construct covariance matrices of the 
the REFLEX II power spectrum and to analyze the possible systematic effects that might affect this measurement. 

Due to the flux-limited selection of the REFLEX II survey, the clustering pattern of galaxy clusters might be affected by 
scale-dependent distortion, as has been observed in galaxy surveys \citep[e.g.][]{teg_sdss,percival_sdss}. Using the mock catalogues, we have shown that these distortions might affect the clustering in configuration space (i.e., when measured with the cluster-correlation function) on scales $r \geq 150\,{\rm Mpc}\, h^{-1}$, which would naively correspond to scales $k\leq 0.04 h {\rm Mpc}^{-1}$ in Fourier space. 
In order to test the impact of this flux-selection effect on the final measurements of power spectrum, we implemented the luminosity dependent estimator of \cite{pvp}, which is designed to correct for this distortion. 
We observed that the shape of the power spectrum measured by means of the FKP estimator does not show significant
distortions compared to the results from the PVP estimator. This implies that the flux-selection of the REFLEX II sample does not introduce a significant systematic effect in
the measurement of the power spectrum of this catalogue.

The shape of the mean power spectrum from our ensemble of mock catalogues 
is in good agreement with the measured power spectrum from the REFLEX II sample, and is statistically distinguishable from the linear perturbation theory predictions on intermediate scales. This implies a clear signature of non-linear evolution in the X-ray cluster spatial distribution. 
Nevertheless, given the level of accuracy of the measurements of power spectrum in the REFLEX II sample, 
it is sufficient to model these distortions using the $Q$-model of \citet{cole2df}.
We find that this prescription provides a good description of the measurements from the mock catalogues
on intermediate scales ($0.02\leq k\,/(h\,{\rm Mpc}^{-1}) \leq 0.25$). This model can also be used to describe the 
shape of the measured REFLEX II power spectrum, providing a valuable tool to extract the cosmological information
contained in the shape of this statistic.
The next generation of X-ray galaxy clusters surveys, such as 
{\it eROSITA}\footnote{http://www.mpe.mpg.de/heg/www/Projects/EROSITA/main.html} and \emph{WFXT}\footnote{http://wfxt.pha.jhu.edu/},
will
provide measurements of the two-point statistics of the cluster population with higher accuracy
than present-day samples, for which a more detailed modelling of
non-linearities will be required \citep[e.g.][]{rpt, montesano}. 

Our measurements of the REFLEX II power spectrum are compatible with the prediction of the $\Lambda$CDM cosmological model and 
shows good agreement with the previous results from the REFLEX sample \citep{reflex1}, save the expected differences due to the lower limiting flux of the REFLEX II sample.
We showed that our measurements cannot provide a statistically significant detection of BAO, which is mainly due to the moderate volume probed by the survey (compared to the volume probed by current galaxy redshift surveys). We found that the power spectra measured from the REFLEX II sample and the mock catalogues are compatible with a scale-independent effective bias in the range of wavenumbers $0.01\leq k\, /(h\,{\rm Mpc}^{-1})\leq 0.1$, 
and that a simple theoretical prediction, based on the halo-mass bias, the halo mass function and
the mass-luminosity relation, is able to describe these measurements. This, together with the modeling of the shape of the power spectrum given by the $Q$-model, provides a link to the cosmological models and allows our measurements to reach their full constraining power.

\section*{Acknowledgments}
We thank Ra\'ul Angulo and Carlton Baugh for providing us with the L-BASICC II simulations.
ABA acknowledges the Ph.D fellowship of the International Max Planck Research School in the OPINAS group at MPE. This research was supported by the DFG cluster of excellence ``Origin and Structure of the Universe''. This publication contains observational data obtained at the ESO La Silla observatory. We wish to thank the ESO Team for their support.


\begin{thebibliography}{}
\bibitem[Angulo et al.(2008)]{angulo} Angulo R., Baugh C. M., Frenk C. S., Lacey G.C.,  2008, MNRAS, 383, 755
\bibitem[Arnaud(1996)]{xspec}Arnaud, K.A., 1996, Astronomical Data Analysis Software and Systems V, eds. Jacoby G. and Barnes J., p17, ASP Conf. Series 101.
\bibitem[Arnaud et al.(2009)]{pressure} Arnaud M., Pratt G. W., Piffaretti R., B\"ohringer H., Croston J.H., Pointecouteau, E., 2009, preprint (astro-ph/0910.1234)
\bibitem[Arnaud et al.(2005)]{arnaud_05} Arnaud M., Pointecouteau E., Pratt G., 2005, A\&A, 441, 893  
\bibitem[Bardeen et al.(1986)]{bardeen} Bardeen J. M., Bond J. R., Kaiser N., Szalay A. S., 1986 ApJ 304, 15
\bibitem[B\"ohringer et al.(2000)]{hb_cga}B\"ohringer H. et al., 2000, ApJS 129, 435
\bibitem[B\"ohringer et al.(2001)]{hb_catalogo}B\"ohringer H. et al., 2001, A\&A, 369, 826 
\bibitem[B\"ohringer et al.(2002)]{reflex_lf}B\"ohringer H. et al., 2002, ApJ, 566, 93
\bibitem[B\"ohringer et al.(2007)]{bohringer_07}B\"ohringer H. et al., 2007, A\&A, 469, 363
\bibitem[Cavaliere \& Fusco-Formiano(1976)]{beta_model}Cavaliere, A., Fusco-Femiano R., 1976, A\&A, 49, 137
\bibitem[Cai et al.(2010)]{cai}Cai Y., Bernstein G., Sheth R. K., preprint (astro-ph/1007.3500)
\bibitem[Cole et al.(2005)]{cole2df} Cole S. et al., 2005, MNRAS, 362, 505
\bibitem[Collins et al.(2000)]{collins_cf} Collins C., et al., 2000, MNRAS, 319, 939
\bibitem[Cooray(2006)]{hmodel_best} Cooray A., 2006, MNRAS, 365, 842
\bibitem[Crocce \& Scoccimarro(2006)]{rpt} Crocce M., Scoccimarro R., 2006, Phys.Rev.D73, 063519
\bibitem[Dickey \& Lockman(1993)]{dickey} Dickey, J.M., Lockman, F.J., 1993, ARA\&A 28, 215
\bibitem[Eisenstein et al.(2005)]{ei} Eisenstein D. et al., 2005, ApJ, 633, 560
\bibitem[Eisenstein \& Hu(1998)]{hu}Eisenstein D., Hu W., 1998, ApJ, 496, 605
\bibitem[Feldman et al.(1994)]{FKP}Feldman H., Kaiser N., Peacock J. A., 1994, ApJ, 426, 23
\bibitem[Frigo \& Johnson(2005)]{fftw}Frigo M., Johnson S., 2005, Proceedings of the IEEE,93, 216
\bibitem[Gazta\~naga et al.(2008)]{gaz}Gazta\~naga E., Cabr\'e A., Hui L., 2008, MNRAS, 399, 1663
\bibitem[Giodini et al.(2009)]{baryon_frac}Giodini S. et al., 2009, ApJ, 703, 982
\bibitem[Guzzo et al.(2009)]{guzzo_optical}Guzzo L. et al., 2009, A\&A, 499, 357
\bibitem[Hockney \& Eastwood(1988)]{hockney}Hockney R. W., Eastwood J. W., 1988, Computer simulation using particles. Bristol, Adam Hilger
\bibitem[H\"utsi(2010)]{hutsi_gc}H\"utsi G., 2010, MNRAS, 401,2477
\bibitem[Jenkins et al.(2001)]{jenkins}Jenkins A., Frenk C., White S. D. M., Colberg J.M., Evrard A.E., Couchman H.M.P., Yoshida N., 2001, MNRAS, 312, 372
\bibitem[Kaiser (1987)]{kaiser}Kaiser N., 1987, MNRAS, 227, 1, 21
\bibitem[Kerscher et al.(2001)]{kerscher}Kerscher M. et al., 2001, A\&A, 377, 1
\bibitem[Komatsu et al.(2010)]{komatsu10}Komatsu E. et al., 2010, preprint (astro-ph/1001.4538)
\bibitem[Lewis \& Bridle (2002)]{lewis}Lewis A., Bridle S., 2002, Phys. Rev. D., 103511
\bibitem[Lukic et al.(2010)]{lukic10}Lukic Z., Reed D.,Habib S., Heitman K., 2010, ApJ, 692, 217
\bibitem[Makino et al(1998)]{makino}Makino N., Sasaki S., Suto Y., 1998, ApJ 497, 555
\bibitem[Mantz et al.(2010)]{mantzI}Mantz A., Allen S., Ebeling H., Rapetti D., Drlica-Wagner A., 2010, MNRAS, in press 
\bibitem[Matarrese et al.(1997)]{matarrese_bi}Matarrese S., Verde L., Heavens A. 1997, MNRAS, 290, 651
\bibitem[Meiksin \& White(1999)]{meiksin}Meiksin A., White M. 1999, MNRAS, 308, 1179
\bibitem[Miller et al.(2001)]{miller}Miller C., Nichol R., Batuski D., 2001, ApJ, 555, 68
\bibitem[Miller \& Batuski (2001)]{miller_batuski}Miller C., Batuski D., 2001, ApJ, 551, 635
\bibitem[Moscardini et al.(2000)]{moscardini}Moscardini L., Matarrese S., Lucchini F., Rosati P.,  2000, MNRAS, 316, 283
\bibitem[Montesano et al.(2010)]{montesano}Montesano F., S\'anchez A. G., Phleps S., 2010, MNRAS, in press 
\bibitem[Padmanabhan et al.(2007)]{padma}Padmanabhan N. et al., 2007, MNRAS, 378, 852
\bibitem[Peebles(1980)]{peebook}Peebles P. J. E., 1980, The Large-Scale structure of the Universe, Princeton University Press, Princeton, NJ
\bibitem[Percival et al.(2002)]{percival02}Percival W. J. et al.,  2002, MNRAS, 337, 1068
\bibitem[Percival et al.(2004)]{pvp}Percival W. J., Verde L., Peacock J. A. 2004, MNRAS, 347, 645
\bibitem[Percival et al.(2007)]{percival_sdss}Percival W. J. et al., 2007, ApJ 657, 645
\bibitem[Percival et al.(2010)]{percival_last}Percival W. J. et al., 2010, MNRAS 401, 2148
\bibitem[Perlmutter et al.(1999)]{perlmutter}Perlmutter S. et al., 1999, ApJ 517, 565
\bibitem[Porciani et al.(1998)]{porciani} Porciani C., Matarrese S., Luccini F., Catelan P., 1998, MNRAS, 298, 1097
\bibitem[Pratt et al.(2009)]{pratt}Pratt G., Croston H. J., Arnaud M., B\"ohringer H., 2009, A\&A 498, 361
\bibitem[Puchwein et al.(2008)]{puchwein}Puchwein E., Sijacki D., Springel V., 2008, ApJ, 687, L53
\bibitem[Press et al.(2002)]{numerical_rec}Press H., Teukolski S., Vetterling W., Flannery B. 2002, Numerical Recipes in C, Cambridge University Press.
\bibitem[Reid et al.(2010)]{reid} Reid B. et al., 2010 MNRAS, 404, 60
\bibitem[Reiprich \& B\"ohringer(1999)]{lm0}Reiprich T., B\"ohringer H., 1999, AN, 320, 296
\bibitem[Rimes \& Hamilton(2006)]{hamilton_rimes}Rimes C., Hamilton A. J. S. 2006, MNRAS, 37, 1205
\bibitem[Riess et al.(2004)]{riess} Riess A.G., Strolger L.G, Tony J. 2004, ApJ 116, 665
\bibitem[S\'anchez et al.(2005)]{ariel_reflexI} S\'anchez A. G., Lambas D. G., B\"ohringer H., Schuecker P., 2005, MNRAS, 362, 1225
\bibitem[S\'anchez et al.(2006)]{sanchez06} S\'anchez A. G. et al., 2006, MNRAS, 366, 189
\bibitem[S\'anchez et al.(2008)]{sanchez_cole} S\'anchez A. G., Cole S. 2008, MNRAS, 385, 830
\bibitem[S\'anchez et al.(2008a)]{ariel1} S\'anchez A. G., Baugh C. M., Angulo R. 2008, MNRAS, 390, 1470
\bibitem[S\'anchez et al.(2009)]{ariel_last}  S\'anchez A. G., Crocce M., Cabr\'e A., Baugh C. M., Gazta\~naga E., 2009, MNRAS, 400, 1643
\bibitem[Schuecker et al.(2001)]{reflex1} Schuecker P. et al.,  2001, A\&A 368, 86
\bibitem[Schuecker et al.(2003)]{reflex1_s8} Schuecker P. et al., 2003, A\&A 398, 867
\bibitem[Smith et al.(2003)]{smith} Smith R. et al., 2003 MNRAS, 341, 1311 
\bibitem[Smith et al.(2007)]{sss}  Smith R., Scoccimarro R., Sheth R. K., 2007, Phys Rev D, 063212
\bibitem[Smith(2008)]{smith_cova}  Smith R., 2008, preprint (astro-ph/0810.1960v2)
\bibitem[Spergel et al.(2007)]{spergel07}  Spergel D. N. et al., 2007 APJSS, 170 
\bibitem[Stanek et al.(2006)]{stanek}  Stanek R., Evrard A. E., B\"ohringer H., Schuecker P., Nord B., 2006, ApJ 209, 17 
\bibitem[Stanek et al.(2010)]{stanek_last}  Stanek R., Rasia E., Evrard A. E., Pierce F., Gazzola L., 2010, ApJ 715, 1508
\bibitem[Suto et al.(2000)]{suto}Suto Y., Magira H., Yamamoto K., 2000, PASJ 52, 249
\bibitem[Takayashi et al.(2009)]{taka1} Takahashi R. et al., 2009, ApJ, 700, 497
\bibitem[Tegmark(1997)]{teg_effv} Tegmark M., 1997, Phys. Rev. Lett 79, 3806.
\bibitem[Tegmark et al. (2006)]{teg_sdss}Tegmark M. et al., 2006, Phys.Rev. D74, 123507
\bibitem[Tinker (2007)]{tinker} Tinker J., 2007, MNRAS, 374, 477
\bibitem[Tinker et al.(2008)]{tinker_b} Tinker J., Kravtsov A., Klypin A., Abazajian K., Warren M., Yepes G., Gottl\"ober S., Holz D., 2008, ApJ, 688, 709
\bibitem[Truemper (1993)]{truemper} Truemper J., 1993, Science 260, no. 5115, 1769
\bibitem[Tsallis(2009)]{tsallis} Tsallis C., 2009, Introduction to Nonextensive Statistical Mechanics, Springer Verlag, NY
\bibitem[Verde \& Heavens(2001)]{verde_tri}Verde L., Heavens A., 2001,  ApJ 553, 14
\bibitem[Vikhlinin et al.(2009)]{vik} Vikhlinin A. et al., 2009, ApJ,  692, 1033-1059 
\bibitem[Wang \& Steinhardt(1998)]{wang}Wang, L., Steinhardt P. J., 1998, ApJ 508, 483 
\bibitem[Warren et al.(2006)]{warren}Warren M., Abazajian K., Holz D., Teodoro L., 2006, ApJ 646, 881
\end{thebibliography}
 \end{document}